\documentclass[aps,prc,twocolumn,showpacs,preprintnumbers,
               nofootinbib,float,superscriptaddress,longbibliography]{revtex4-1}
\usepackage{graphicx}
\usepackage{color}
\usepackage[dvipsnames]{xcolor}
\usepackage[colorlinks=true, pdfstartview=FitV, linkcolor=RedOrange, citecolor=ForestGreen, urlcolor=blue]{hyperref}
\usepackage{amsmath}
\usepackage{amssymb}
\usepackage{bm}
\usepackage{physics}
\usepackage{placeins}

\newcommand{\bs}{\boldsymbol}

\usepackage[maxfloats=256]{morefloats}
\begin{document}
\author{Hendrik Roch} \email{Hendrik.Roch@wayne.edu}
\affiliation{Department of Physics and Astronomy, Wayne State University, Detroit, Michigan 48201, USA}
\author{Syed Afrid Jahan} \email{hm0746@wayne.edu}
\affiliation{Department of Physics and Astronomy, Wayne State University, Detroit, Michigan 48201, USA}
\author{Chun Shen} \email{chunshen@wayne.edu}
\affiliation{Department of Physics and Astronomy, Wayne State University, Detroit, Michigan 48201, USA}

\title{On model emulation and closure tests for 3+1D relativistic heavy-ion collisions}

\begin{abstract}
In nuclear and particle physics, reconciling sophisticated simulations with experimental data is vital for understanding complex systems like the Quark Gluon Plasma (QGP) generated in heavy-ion collisions.
However, computational demands pose challenges, motivating using Gaussian Process emulators for efficient parameter extraction via Bayesian calibration.
We conduct a comparative analysis of Gaussian Process emulators in heavy-ion physics to identify the most adept emulator for parameter extraction with minimal uncertainty.
Our study contributes to advancing computational techniques in heavy-ion physics, enhancing our ability to interpret experimental data and understand QGP properties.
\end{abstract}

\maketitle

\section{Introduction}
\label{s:intro}
In nuclear and particle physics, comparing complex simulations with multiple experimental observables is important for elucidating the properties of systems under study. 
A quintessential example lies in nuclear physics, which investigates the Quark Gluon Plasma (QGP) via relativistic heavy-ion collisions. 
Experiments at the Relativistic Heavy Ion Collider (RHIC) and the Large Hadron Collider (LHC) aim to recreate the extreme conditions akin to those prevailing in the early Universe shortly after the Big Bang~\cite{Busza:2018rrf}.

However, direct measurement of the resultant hot and dense QGP state remains elusive due to its transient existence and confinement to small spatial dimensions within collider experiments. 
Understanding the QGP's properties necessitates analyzing experimental measurements, which imprint the entire collision history as the QGP is created, expands, and fragments into a collection of hadrons~\cite{Busza:2018rrf}. 
Theoretical frameworks for modeling such collisions have undergone substantial advancement over recent decades, encompassing various stages from the initial state~\cite{Miller:2007ri,Loizides:2014vua,Broniowski:2007nz,Drescher:2006ca,Schenke:2012wb,Paatelainen:2012at,Moreland:2014oya,Giacalone:2019kgg,Shen:2017bsr} to hadronic rescattering dynamics~\cite{Bass:1998ca,SMASH:2016zqf,Cassing:1999es,Lin:2004en}, including pre-equilibrium~\cite{Schlichting:2019abc}, relativistic viscous hydrodynamics~\cite{Teaney:2009qa,Romatschke:2017ejr}, and particlization~\cite{Cooper:1974mv}.

These models, containing numerous constituent elements, demand careful calibration against experimental data, entailing the tuning of many parameters simultaneously to extract physics from the observations. 
Moreover, the computational demands pose a substantial challenge, with long simulation runtime required even for a single event simulation. 
By inferring from experimental data, this model parameter determination embodies an inverse problem paradigm.

Bayesian inference, leveraging Bayes's theorem, offers a principled approach to address inverse problems by discerning parameter sets and their associated uncertainties systematically~\cite{Sivia2006}. 
However, the exhaustive evaluation of models across diverse parameter spaces often proves infeasible due to computational constraints.
Herein lies the utility of model emulation, which, once trained with model outputs, furnishes computationally inexpensive surrogates for the full model.
These emulators, by virtue of their efficiency, replace the full model in Bayesian analyses, thereby facilitating parameter calibration.

This study compares various Gaussian Process (GP) emulators for (3+1)D simulations of relativistic heavy-ion physics in the RHIC Beam Energy Scan program. 
Our objective is to discern the most adept GP emulator for a given training dataset, characterized by minimal prediction uncertainty, for parameter extraction via Bayesian calibration. 
Additionally, we aim to quantify the influence of GP emulators with different accuracies on the model parameter extraction through closure tests.

In Sec.~\ref{sec:Bayes}, we concisely discuss the requisite components for the Bayesian inference approach for inverse problems. 
Following this, Sec.~\ref{sec:GP} presents fundamental theories for training the GP emulators. 
Subsequently, in Sec.~\ref{sec:GP_comparison}, we compare three open-source GP emulator implementations, where we look at the uncertainty and quality of their prediction in different scenarios.
Then, Section~\ref{sec:MCMC} analyzes the posterior distributions obtained from these different GPs in a closure test. We also perform a local sensitivity analysis for the observables under consideration in Sec.~\ref{sub:obs_sensitivity_analysis}.
Finally, we will conclude and give an outlook in Sec.~\ref{sec:conclusion_outlook}.

\section{Bayesian Inference}
\label{sec:Bayes}

This section provides an overview of the application of Bayesian Inference with multi-stage Monte Carlo event simulations in relativistic heavy-ion collisions, highlighting their utility in mitigating the computational costs associated with such simulations.

Multi-stage event-by-event simulation frameworks, such as iEBE-VISHNU~\cite{Shen:2014vra, Bernhard:2016tnd}, JETSCAPE~\cite{Putschke:2019yrg}, Trajectum~\cite{Nijs:2020roc}, and iEBE-MUSIC~\cite{Shen:2022oyg}, are essential tools for studying various physics phenomena in relativistic heavy-ion collisions.
However, these simulations are often computationally intensive.
To alleviate this computational burden, emulators are employed to predict the model output for a given parameter set~\cite{Ohagan2001, Higdon2004}.

Before delving into the specifics of the emulators compared in this study, we introduce the concept of Bayesian model calibration.
Let $\bs{\theta} \equiv (\theta_1, \dots, \theta_m)$ denote the parameters of the model, and let ${\bf y}_{\rm sim}(\cdot)$ represent the model function.
This function maps the parameter set $\bs{\theta}$ to simulation outputs in $\mathbb{R}^d$, where $d$ is the dimension of the observable space.
These simulation outputs are then compared to experimental values, denoted by ${\bf y}_{\rm exp} \equiv (y_{\rm exp, 1}, \dots, y_{{\rm exp}, d})$.
Thus, a statistical model can be formulated as:
\begin{equation}
{\bf y}_{\rm exp} = {\bf y}_{\rm sim}(\bs{\theta}) + \bs{\epsilon},
\label{eq:model}
\end{equation}
where $\bs{\epsilon}$ represents the residual error following a multivariate normal (MVN) distribution with mean $\mathbf{0}$ and covariance matrix $\bs{\Sigma}$.

The problem at hand is inherently an inverse problem, where we seek to identify the parameters $\bs{\theta}$ that best describe the observed data.
To address this, we adopt a Bayesian approach to inference, where probability is interpreted as the degree of belief accounting for all available information~\cite{Sivia2006}.
Bayes' rule provides a framework for updating our beliefs based on observed data:
\begin{equation}
\mathcal{P}(\bs{\theta} | {\bf y}_{\rm exp}) = \frac{\mathcal{P}({\bf y}_{\rm exp} | \bs{\theta}) \mathcal{P}(\bs{\theta})}{\mathcal{P}({\bf y}_{\rm exp})}.
\label{eq:bayes}
\end{equation}
Here, $\mathcal{P}(\bs{\theta} | {\bf y}_{\rm exp})$ represents the posterior probability density function, $\mathcal{P}(\bs{\theta})$ is the prior probability density function, $\mathcal{P}({\bf y}_{\rm exp})$ stands for the total evidence, and $\mathcal{P}({\bf y}_{\rm exp} | \bs{\theta})$ is the likelihood function, which, given the MVN distribution of $\bs{\epsilon}$, can be expressed as:
\begin{equation}
\frac{1}{\sqrt{|2\pi\bs{\Sigma}|}}\exp\left[-\frac{1}{2}({\bf y}_{\rm sim}(\bs{\theta})-{\bf y}_{\rm exp})^\mathsf{T}\bs{\Sigma}^{-1}({\bf y}_{\rm sim}(\bs{\theta})-{\bf y}_{\rm exp})\right].
\label{eq:likelihood}
\end{equation}

To obtain the posterior distribution, which is often intractable analytically, Markov Chain Monte Carlo (MCMC) techniques are employed~\cite{brooks2011handbook, Trotta2008}.
These methods enable the estimation of posterior distributions for $\bs{\theta}$ and quantifying their uncertainties.

In practice, evaluating the right-hand side of Eq.~\eqref{eq:bayes} for numerous $\bs{\theta}$ sets in MCMC simulations can be computationally prohibitive.
Here, GP emulators offer a solution by providing computationally efficient surrogates for the expensive simulations~\cite{gramacy2020surrogates, Rasmussen2004}.
These emulators are trained using simulation data for various parameter sets $\bs{\theta}$.
Once trained, the emulators can predict the mean $\bs{\mu}(\bs{\theta})$ and covariance ${\bf C}(\bs{\theta})$ of the simulation output ${\bf y}_{\rm exp}(\bs{\theta})$.
Subsequently, the likelihood function (Eq.~\eqref{eq:likelihood}) can be approximated as~\cite{Ohagan2001}:
\begin{equation}
\frac{1}{\sqrt{|2\pi{\bf V}(\bs{\theta})|}}\exp\left[-\frac{1}{2}(\bs{\mu}(\bs{\theta})-{\bf y}_{\rm exp})^\mathsf{T}{\bf V}(\bs{\theta})^{-1}(\bs{\mu}(\bs{\theta})-{\bf y}_{\rm exp})\right],
\label{eq:approx_likelihood}
\end{equation}
where ${\bf V}(\bs{\theta})={\bf C}(\bs{\theta})+\bs{\Sigma}$.
This approximation facilitates the MCMC sampling process, enabling efficient exploration of the posterior probability distribution function.

\section{Gaussian Process Emulators}
\label{sec:GP}

This section presents a comparative analysis of different implementations of GP emulators employed in our study.
Specifically, we investigate two GP emulator methodologies, namely the PCGP and the PCSK methods, which are integrated into the Python package \texttt{surmise} by the BAND collaboration~\cite{surmise2023}.
Additionally, we consider the standard GP emulator available in the Scikit-learn Python package. The Scikit GP was used in a few previous Bayesian analyses in our field~\cite{Bernhard:2019bmu, JETSCAPE:2020mzn, Nijs:2020ors, Parkkila:2021tqq, Mantysaari:2022ffw, Heffernan:2023gye, Soeder:2023vdn, Shen:2023awv, Shen:2023pgb}.

All these emulators share a basis vector approach~\cite{Higdon2008}, tailored to address the challenges posed by the high-dimensional output space of the model. 
Principal Component Analysis (PCA)~\cite{Ramsay97functionaldata} is utilized to project the high-dimensional output onto a lower-dimensional space, enabling efficient emulation.
Specifically, each output for a principal component is independently trained by a GP emulator.

We begin by elucidating the fundamental concept behind GP emulation with PCA.
Let $\lbrace\bs{\theta}_1^{\text{tr}}, \dots, \bs{\theta}_n^{\text{tr}}\rbrace$ represent a training parameter set comprising $n$ unique training points for the emulators.
Each training point $\bs{\theta}_i^{\text{tr}}$ generates a $d$-dimensional vector $\bar{\mathbf{y}}_{\text{sim}}(\bs{\theta}_i^{\text{tr}})$, where $d$ denotes the number of observables. The model simulation results $\bar{\mathbf{y}}_{\text{sim}}(\bs{\theta}_i^{\text{tr}})$ are averaged over multiple collision events.
Before applying the PCA transformation, we standardize the training data as, $\tilde{\bf y}_{\rm sim}(\bs\theta_i^{\rm tr}) \equiv \mathbf{G}^{-1} (\bar{\bf y}_{\rm sim}(\bs\theta_i^{\rm tr})-\bf h)$, using a centering vector $\mathbf{h}$ and scaling matrix $\mathbf{G} = \mathrm{diag}(\bf c)$.
The resulting $\tilde{\bf y}_{\rm sim}(\bs\theta_i^{\rm tr})$ has a zero mean and unity variance over the $n$ training samples.
We define a $d \times n$ matrix $\mathbf{\Xi} \equiv \{ \tilde{\mathbf{y}}_{\text{sim}}(\bs{\theta}_1^{\text{tr}}), \dots, \tilde{\mathbf{y}}_{\text{sim}}(\bs{\theta}_n^{\text{tr}})\}$.
PCA yields a linear transformation matrix ${\bf S}=\left[{\bf s}_1,\dots,{\bf s}_p\right]$ of dimension $d \times p$, which can project the output matrix $\bs\Xi$ from a $d$-dimensional space to a $p$-dimensional principal component space via the transformation ${\bf T}=\left[{\bf t}_1,\dots,{\bf t}_p\right]=\bf S^{\mathsf{T}}\bs\Xi$.
In most applications, keeping only $p \ll d$ principal components to describe almost all the variance in the original dataset is sufficient~\cite{Higdon:2014tva,Surer:2022lhs,Liyanage:2023nds,doi:10.1080/00401706.2023.2210170}.

Subsequently, an independent GP emulator is constructed for each of the outputs $\mathbf{t}_l(\bs{\theta})={\bf s}_l^\mathsf{T} \tilde{\bf y}_{\rm sim}(\bs\theta_i^{\rm tr})$.
The emulator predicts the mean $m_l(\bs{\theta})$ and variance $s^2_l(\bs{\theta})$ for each principal component $t_l$ with $l = 1, \dots, p$, given by
\begin{equation}
t_l(\bs{\theta}) | \bs{t}_l \approx \text{N}(m_l(\bs{\theta}), s^2_l(\bs{\theta})),
\end{equation}
where $m_l(\bs{\theta}) = \mathbf{k}_l^\mathsf{T}\mathbf{K}_l^{-1}\mathbf{t}_l$ and $s^2_l(\bs{\theta}) = k_l(\bs{\theta}, \bs{\theta}) - \mathbf{k}_l^\mathsf{T}(\bs{\theta})\mathbf{K}_l^{-1}\mathbf{k}_l(\bs{\theta})$. 
Here, $\mathbf{k}_l(\bs{\theta})=\left[k_l(\bs\theta,\bs\theta^{\rm tr}_i)\right]^n_{i=1}$ denotes the covariance vector between $n$ training points $\lbrace\bs\theta_1^{\rm tr},\dots,\bs\theta_n^{\rm tr}\rbrace$ and any chosen parameter vector $\bs{\theta}$, and $\mathbf{K}_l = \left[k_l(\bs\theta_i^{\rm tr},\bs\theta_j^{\rm tr}) + \delta_{i, j}r_{l, i}\right]_{i, j = 1}^n$ represents the covariance matrix between the $n$ training points. 
The covariance function $k_l(\bs{\theta}, \bs{\theta}')$ depends on the choice of the kernel function in the GP.\footnote{The PCSK and PCGP emulators use a Matérn kernel (smoothness parameter $3/2$), while we choose to use a Radial Basis Function (RBF) kernel for the Scikit GP. The Scikit GP setup was used in a few of the previous Bayesian analyses~\cite{JETSCAPE:2020mzn, Shen:2023awv}.}
Additionally, the term $\delta_{i,j}r_{l, i}$, along with the Kronecker delta, accounts for uncertainty arising from stochastic simulation results for the different events.
Here $r_{l, i}$ is the square of the statistical uncertainty of the $l$-th principal component at the training point $\bs{\theta}_i$.

The term $\delta_{i,j}r_{l, i}$ is the main difference between the PCGP/Scikit GP implementations of GPs used in almost all heavy-ion physics related studies~\cite{Petersen:2010zt,Novak:2013bqa,Sangaline:2015isa,Bernhard:2015hxa,Bernhard:2016tnd,Moreland:2018gsh,Bernhard:2019bmu,JETSCAPE:2020shq,Nijs:2020ors,Nijs:2020roc,JETSCAPE:2020mzn,Heffernan:2023utr,Heffernan:2023gye} and the PCSK method used in~\cite{Liyanage:2023nds}. 
Instead of only taking into account the mean values of the training data, the PCSK method also considers the standard deviation at the training points.
This is particularly relevant as simulations of heavy-ion collisions with different initial states for given parameters $\bs\theta$ yield observables with a certain variation.
The PCSK emulation incorporates this variation in the accuracy between the resulting data points, known as heteroskedasticity, by using stochastic kriging~\cite{Ankenman2010}.
We explore the effects of the size of statistical errors on the trained emulator accuracy in Appendix~\ref{app:PCSK_err}.

After training, the emulators predict the mean $m_l(\bs\theta)$ and variance $s^2_l(\bs\theta)$ for each observable at a given point $\bs{\theta}$. 
The outputs for each of the GP emulators $l=1,\dots,p$ are transformed back to the original $d$-dimensional space using the inverse PCA transformation.
Thus, the $p$-dimensional emulator prediction mean $\bs{m}(\bs\theta)=(m_1(\bs\theta),\dots,m_p(\bs\theta))$ and a $p\times p$ covariance matrix $\bs\sigma(\bs\theta)$ with diagonal elements $s^2_l(\bs\theta)$ is used, such that the inverse PCA transformation yields
\begin{equation}
    \bs{\mathcal{G}}(\bs{\theta})\approx \mathrm{MVN}(\bs\mu(\bs\theta),{\bf C}(\bs\theta)),
\end{equation}
where $\bs\mu(\bs\theta)={\bf h}+{\bf G}{\bf S}\bs{m}(\bs\theta)$ and ${\bf C}(\bs\theta)={\bf G}{\bf S}{\bs\sigma(\bs\theta)}{\bf S}^{\mathsf{T}}{\bf G}$ are the emulator prediction mean and covariance respectively.
The likelihood at a parameter space point $\bs{\theta}$ is obtained using Eq.~\eqref{eq:approx_likelihood}, and Eq.~\eqref{eq:bayes} is subsequently utilized to derive the posterior probability distribution $\mathcal{P}(\bs{\theta}|\mathbf{y}_{\text{exp}})$.

\section{GP Emulator Comparison}
\label{sec:GP_comparison}
In Sec.~\ref{sub:emu_setup} we will describe our training setups in this work, additionally we are going to introduce a PCA transformation for functional model parameters in Sec.~\ref{sub:PCA_trafo_theory} and the characteristic measures for the emulator precision in Sec.~\ref{sub:test_metrics}.
We will discuss the validation of the GP emulators for different scenarios in Sec.~\ref{sub:emu_validation}.

\subsection{Training Setup}
\label{sub:emu_setup}

In this work, we divide the experimental observables into several groups and apply the principal component Gaussian Process Regression for each group~\cite{Liyanage:2023nds, Shen:2023awv}.
In our work, we train GP emulators for different observable groups first separated by the system's collision energy.
We further split the particle multiplicities observables from the anisotropic flow coefficients and average transverse momenta because we will test the GP emulators' accuracy for training with multiplicities or the logarithm of the data.
The logarithmic transformation was adopted by a few previous works~\cite{JETSCAPE:2020mzn, Nijs:2020roc}.

Table~\ref{tab:training_data} shows the different observables and the number of data points we grouped for the training in the GP emulators for this analysis used before in Ref.~\cite{Shen:2023awv}.
\begin{table}[tb!]
    \caption{The training data from Au+Au collisions used in this study. The datasets are separated in a way, which is later on used for the training of the emulators. The last column shows the number of data points for the training set.}
    \label{tab:training_data}
    \centering
    \begin{tabular}{c|c|c}
        \hline\hline
        Training Set & Observables & Points \\
        \hline
        1 & $\mathrm{d}N/\mathrm{d}y|_{\pi^+,K^+,p}^{7.7\;\rm{GeV}}$ & 21 \\
        2 & $\langle p_{\rm T}\rangle|_{\pi^+,K^+,p,\bar{p}}^{7.7\;\rm{GeV}}$, $v_{2/3}^{\rm ch}\lbrace 2\rbrace|^{7.7\;\rm{GeV}}$ & 40 \\
        3 & $\mathrm{d}N/\mathrm{d}y|_{\pi^+,K^+,p}^{19.6\;\rm{GeV}}$ & 21 \\
        4 & $\langle p_{\rm T}\rangle|_{\pi^+,K^+,p,\bar{p}}^{19.6\;\rm{GeV}}$, $v_{2/3}^{\rm ch}\lbrace 2\rbrace|^{19.6\;\rm{GeV}}$ & 40 \\
        5 & $\mathrm{d}N/\mathrm{d}y|_{\pi^+,K^+,p,\bar{p}}^{200\;\rm{GeV}}$ & 28 \\
        6 & $\langle p_{\rm T}\rangle|_{\pi^+,K^+,p,\bar{p}}^{200\;\rm{GeV}}$, $v_{2/3}^{\rm ch}\lbrace 2\rbrace|^{200\;\rm{GeV}}$ & 40 \\
        7 & $\mathrm{d}N_{\rm ch}/\mathrm{d}\eta|^{19.6\;\rm{GeV}}$ & 170 \\
        8 & $\mathrm{d}N_{\rm ch}/\mathrm{d}\eta|^{200\;\rm{GeV}}$ & 170 \\
        9 & $v_2^{\rm ch}(\eta)|^{200\;\rm{GeV}}$ & 14 \\
        \hline\hline
    \end{tabular}
\end{table}
The datasets contain event-by-event simulations of Au+Au collisions at three energies: 200, 19.6, and 7.7 GeV.
The training dataset contains 1000 design points determined with a Maximum Projection Latin Hypercube Design algorithm~\cite{doi:10.1080/00224065.2019.1611351,10.1093/biomet/asv002} to explore the $\bs\theta$-space.
This algorithm orders the design points to ensure uniform coverage of model parameter space with a subset of design points.
This feature allows us to quantify how the emulator accuracy scales with the size of training datasets.
In the following, we denote the design points from this algorithm by ``LHD points''. 
For each design point, 1000 minimum bias events were simulated at 200 GeV and 2000 minimum bias events at 19.6 and 7.7 GeV each.\footnote{Points with relative statistical uncertainty larger than 10\% are discarded before training the GP emulators.}
Additionally, 100 parameter sets were drawn from the resulting posterior distributions in Ref.~\cite{Shen:2023awv}, and simulations were carried out for these points.
This allows us to study, in an active learning scenario, whether the uncertainty in the emulators can be decreased when the density of training points is increased in regions where the likelihood of the final parameters is more significant.

The simulations were performed with a 20-dimensional parameter space, summarized in Table~\ref{tab:parameters} with the ranges of the uniform priors.
Their definitions and the physics behind them can be found in Ref.~\cite{Shen:2023awv}.
In this work, we want to focus on the accuracy of the GP emulators and their impact on Bayesian parameter extraction in closure tests.
\begin{table}[tb!]
    \caption{The 20 model parameters and their prior ranges.}
    \label{tab:parameters}
    \centering
    \begin{tabular}{c|c|c|c}
    \hline\hline
    Parameter & Prior & Parameter & Prior \\
    \hline
    $B_G\;[\mathrm{GeV}^{-2}]$ & $[1,25]$ & $\alpha_{\text{string tilt}}$ & $[0,1]$ \\
    $\alpha_{\rm shadowing}$ & $[0,1]$ & $\alpha_{\text{preFlow}}$ & $[0,2]$ \\
    $y_{{\rm loss},2}$ & $[0,2]$ & $\eta_0$ & $[0.001,0.3]$ \\
    $y_{{\rm loss},4}$ & $[1,3]$ & $\eta_2$ & $[0.001,0.3]$ \\
    $y_{{\rm loss},6}$ & $[1,4]$ & $\eta_4$ & $[0.001,0.3]$ \\
    $\sigma_{y_{\rm loss}}$ & $[0.1,0.8]$ & $\zeta_{\rm max}$ & $[0,0.2]$ \\
    $\alpha_{\rm rem}$ & $[0,1]$ & $T_{\zeta,0}\;[{\rm GeV}]$ & $[0.15,0.25]$\\
    $\lambda_B$ & $[0,1]$ & $\sigma_{\zeta,+}\;[{\rm GeV}]$ & $[0.01,0.15]$ \\
    $\sigma_x^{\rm string}\;[{\rm fm}]$ & $[0.1,0.8]$ & $\sigma_{\zeta,-}\;[{\rm GeV}]$ & $[0.005,0.1]$ \\
    $\sigma_\eta^{\rm string}$ & $[0.1,1]$ & $e_{\rm sw}\;[{\rm GeV}/{\rm fm}^3]$ & $[0.15,0.5]$\\
    \hline\hline
    \end{tabular}
\end{table}

\subsection{PCA Transformation for Functional Model Parameters}
\label{sub:PCA_trafo_theory}
Our model simulations take in functional inputs for some key physics quantities of interest, namely the average initial-state rapidity loss $\langle y_{\rm loss}\rangle (y_\mathrm{init})$ and QGP specific shear and bulk viscosities $\tilde{\eta}(\mu_B)$ and $\tilde{\zeta}(T, \mu_B)$~\cite{Shen:2023awv}.
The average initial-state rapidity loss for each parton pair is parametrized as a piece-wise function of the incoming rapidity $y_{\rm init}$ in the collision pair rest frame,
\begin{equation}
    \langle y_{\rm loss}\rangle = \begin{cases}
        y_{\rm loss,2}\frac{y_{\rm init}}{2},\; 0<y_{\rm init}\leq 2\\
        y_{\rm loss,2} + (y_{\rm loss,4} - y_{\rm loss,2})\frac{y_{\rm init}-2}{2},\; 2<y_{\rm init}<4\\
        y_{\rm loss,4} + (y_{\rm loss,6} - y_{\rm loss,4})\frac{y_{\rm init}-4}{2},\; y_{\rm init}\geq 4
    \end{cases}
    \label{eq:ylossParam}
\end{equation}
with the parameter $y_{{\rm loss}, n}$ being the average amount of rapidity loss for $y_{\rm init}=n$.
The $\mu_B$-dependence of the QGP's specific shear viscosity is parametrized by
\begin{equation}
    \tilde{\eta}(\mu_B)=\begin{cases}
        \eta_0 + (\eta_2 - \eta_0)\frac{\mu_B}{0.2},\; 0<\mu_B\leq 0.2\;\mathrm{GeV}\\
        \eta_2 + (\eta_4 - \eta_2)\frac{(\mu_B - 0.2)}{0.2},\; 0.2<\mu_B<0.4\;\mathrm{GeV}\\
        \eta_4,\; \mu_B\geq 0.4\;\mathrm{GeV}
    \end{cases}
    \label{eq:QGPshearParam}
\end{equation}
where $\tilde{\eta}=\eta T/(e+P)$ with the parameters $\eta_0$, $\eta_2$, $\eta_4$ are the values of the specific shear viscosity at $\mu_B=0,0.2,0.4\;\mathrm{GeV}$ respectively.
The functional dependence for the QGP's specific bulk viscosity $\tilde{\zeta}$ is assumed to be an asymmetric Gaussian of the form
\begin{equation}
    \tilde{\zeta}(T,\mu_B)=\begin{cases}
        \zeta_{\rm max}\exp\left[-\frac{(T-T_\zeta(\mu_B))^2}{2\sigma^2_{\zeta,-}}\right],\; T<T_\zeta(\mu_B) \\
        \zeta_{\rm max}\exp\left[-\frac{(T-T_\zeta(\mu_B))^2}{2\sigma^2_{\zeta,+}}\right],\; T\geq T_\zeta(\mu_B)
    \end{cases},
    \label{eq:QGPbulkParam}
\end{equation}
where $\tilde{\zeta}\equiv \zeta T/(e+P)$ with the bulk pressure peak temperature $T_\zeta(\mu_B)\equiv T_{\zeta,0}-\frac{0.15}{1\;\mathrm{GeV}}\mu_B^2$.

Although the model parameters in Table~\ref{tab:parameters} provide plenty of information to the GP emulators, the functional form of the parametrizations is not included in the training.
One way to provide the GP emulators with this information is to enlarge the dimension of the model parameter space and feed the GP emulators with more sampled points for $\langle y_{\rm loss}\rangle (y_\mathrm{init})$ and QGP specific shear and bulk viscosities $\tilde{\eta}(\mu_B)$ and $\tilde{\zeta}(T, \mu_B)$.
This approach was carried out in the previous JETSCAPE analysis and effectively increased emulator accuracy~\cite{JETSCAPE:2020mzn, JETSCAPE:2020shq}.
Here, we introduce a more systematic way to deal with functional model parameters using PCA transformation.

We use the parametrizations in Eqs.~\eqref{eq:ylossParam}-\eqref{eq:QGPbulkParam} to generate the corresponding functions at each training point $\bs{\theta}_i^{\text{tr}}$ with 100 equally spaced points in the range $T\in\left[0,0.5\right]\;\mathrm{GeV}$ for $\tilde{\zeta}$, $\mu_B\in\left[0,0.6\right]\;\mathrm{GeV}$ for $\tilde{\eta}$ and $y_{\rm init}\in\left[0,6.2\right]$ for $\langle y_{\rm loss}\rangle$.
Then, we perform the standardization and PCA similar to the one described in Sec.~\ref{sec:GP}.
We proceed with as many principal components as necessary to explain 99 percent of the variance of each parametrization and substitute the four parameters $\zeta_{\rm max}$, $T_{\zeta,0}$, $\sigma_{\zeta,+}$ and $\sigma_{\zeta,+}$ by the needed number of principal components.
Similarly, we substitute the $\eta_0$, $\eta_2$, $\eta_4$ parameters for the shear viscosity and the three rapidity loss parameters $y_{\rm loss,2}$, $y_{\rm loss,4}$ and $y_{\rm loss,6}$.\footnote{The bulk viscosity parameterized in Eq.~\eqref{eq:QGPbulkParam} needs six principal components, while three components are sufficient in the QGP specific shear viscosity in Eq.~\eqref{eq:QGPshearParam} and averaged initial-state rapidity loss in Eq.~\eqref{eq:ylossParam}.}
The emulator is then trained with the principal components instead of the parameters of the parametrization.
When the prediction with a parameter set $\bs{\theta}$ is performed, the original parameters are then transformed into the PCA space with the same transformation, such that the transformation is completely encapsulated in the emulator.

\subsection{Test Metrics for Model Emulators}
\label{sub:test_metrics}
To quantify the prediction error of the GP emulators, we define
\begin{equation}
    \mathcal{E}\equiv\sqrt{\left\langle\left(\frac{\text{prediction}-\text{truth}}{\text{truth}}\right)^2\right\rangle}
    \label{eq:uncertainty}
\end{equation}
for each observable in the analysis.
The GP emulator provides the covariance matrix at the prediction point, which is used as an estimation for the prediction uncertainty.
To test how reliable the GP emulators' uncertainty is, we define
\begin{equation}
    \mathcal{H}\equiv\ln\left(\sqrt{\left\langle\left(\frac{\text{prediction}-\text{truth}}{\text{prediction uncertainty}}\right)^2\right\rangle}\right).
    \label{eq:honesty}
\end{equation}
For an accurate prediction, we expect the values of $\mathcal{E} \rightarrow 0$ and $\mathcal{H} \rightarrow 0$.
In the case where $\mathcal{H}>0$, the emulator gives uncertainties that are too small compared to the actual error away from the true values; when $\mathcal{H}<0$, the returned uncertainty estimates are too conservative.

\subsection{Emulator Validation}
\label{sub:emu_validation}
We will now validate the results of the three different emulators by investigating $\mathcal{E}$ and $\mathcal{H}$ in various setups.
First, we will look at the dependence in terms of the number of LHD training points (Sec.~\ref{subsub:train_point_dependence}), and then consider the values including most of the LHD points; we compare the performance of the three types of the emulator in Sec.~\ref{sub:emu_uncertainties}.
We further will investigate the ansatz with the logarithmic training for the multiplicity observables (Sec.~\ref{subsub:log_training}), followed by the inclusion of additional PCA transformations for parametrized curves (Sec.~\ref{subsub:PCA_trafo}).
Lastly, we will discuss the active learning approach by including 70 additional points from the posterior distribution (Sec.~\ref{subsub:active_learning}).

\subsubsection{Training Size Dependence for Model Emulation Performance}
\label{subsub:train_point_dependence}
Before we consider applying all design points for the training of the emulators, we will first investigate how the emulation accuracy (Eq.~\eqref{eq:uncertainty}) and the quality of uncertainty estimation (Eq.~\eqref{eq:honesty}) scale with the size of training points.
For this analysis, we will only show three examples and focus on the general qualitative behavior of the emulators when the training set varies in size.

\begin{figure*}
    \centering
    \includegraphics[width=0.49\textwidth]{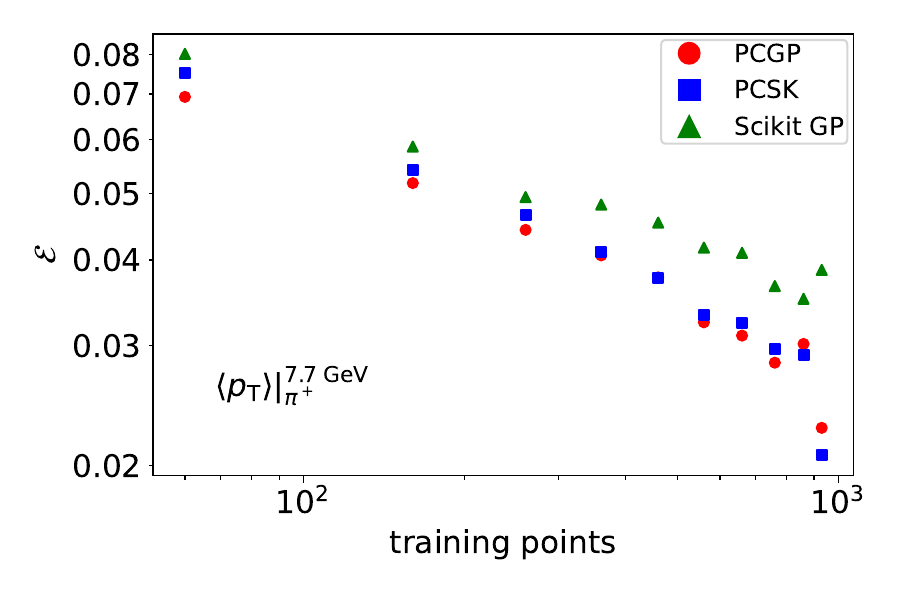}
    \includegraphics[width=0.49\textwidth]{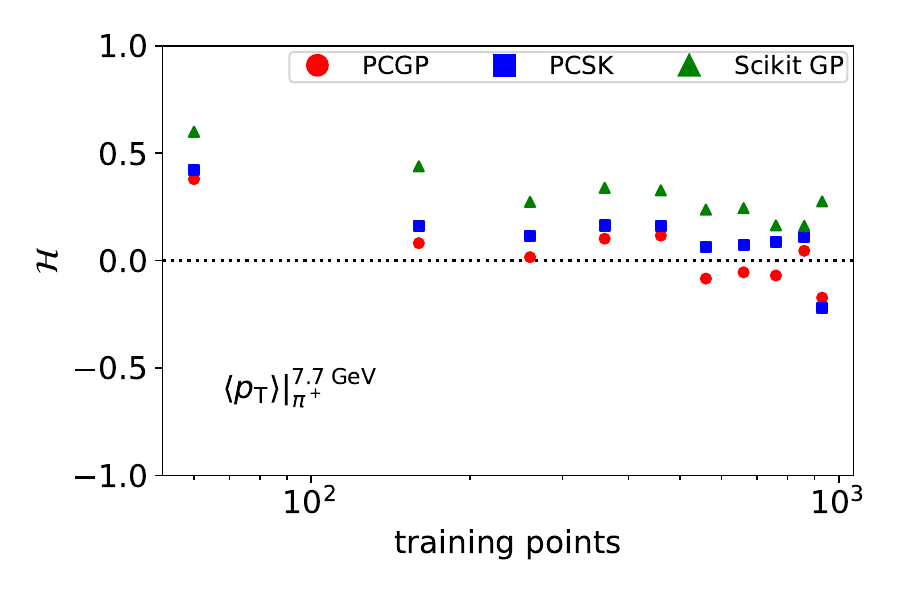}
    \includegraphics[width=0.49\textwidth]{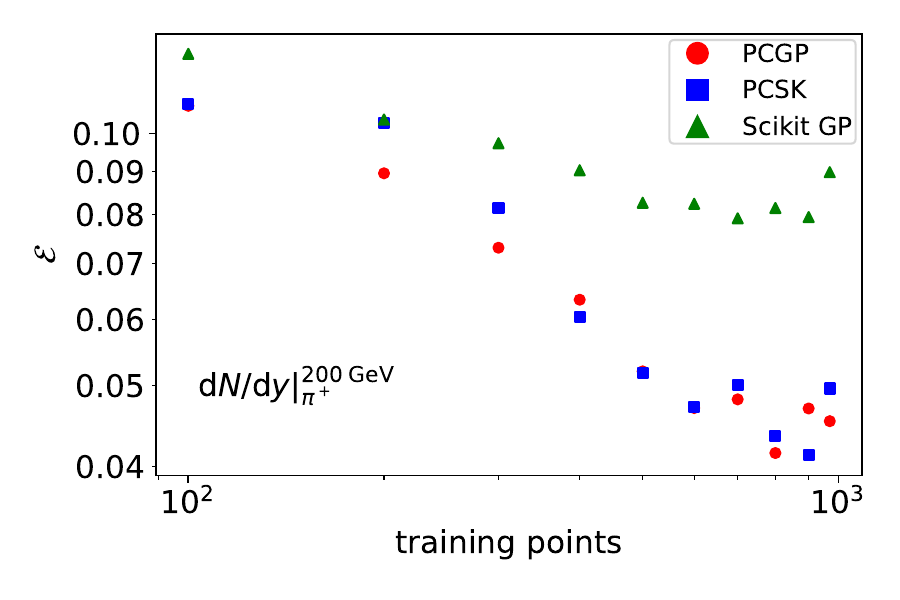}
    \includegraphics[width=0.49\textwidth]{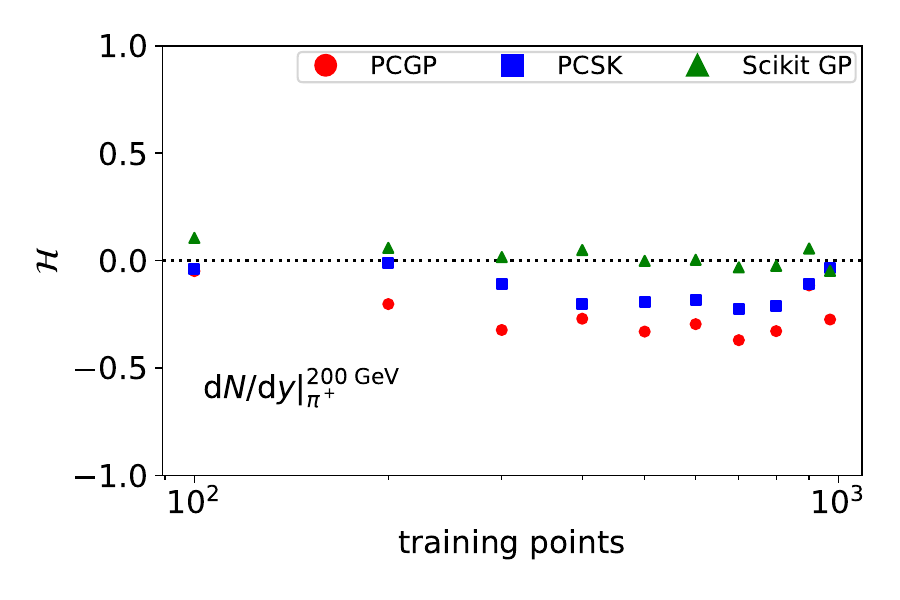}
    \includegraphics[width=0.49\textwidth]{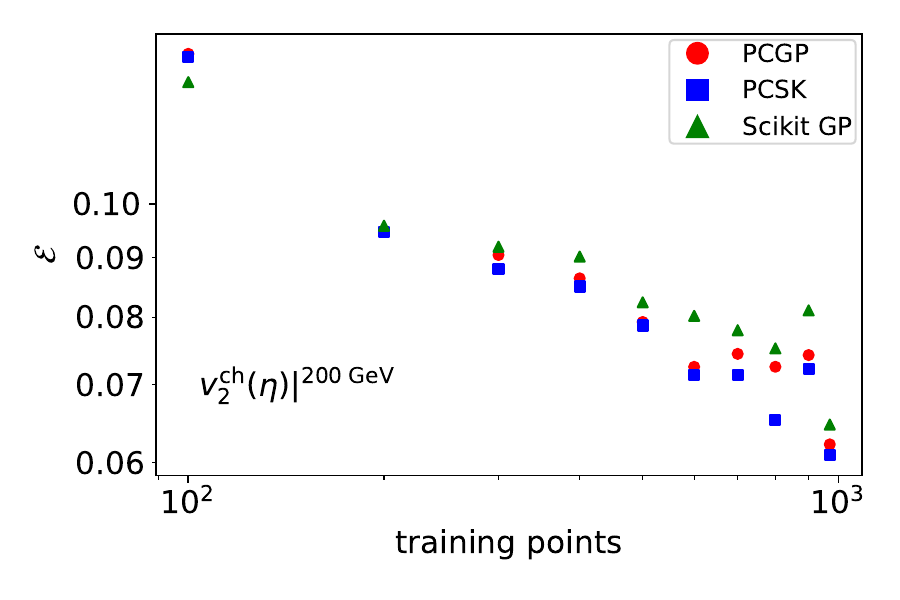}
    \includegraphics[width=0.49\textwidth]{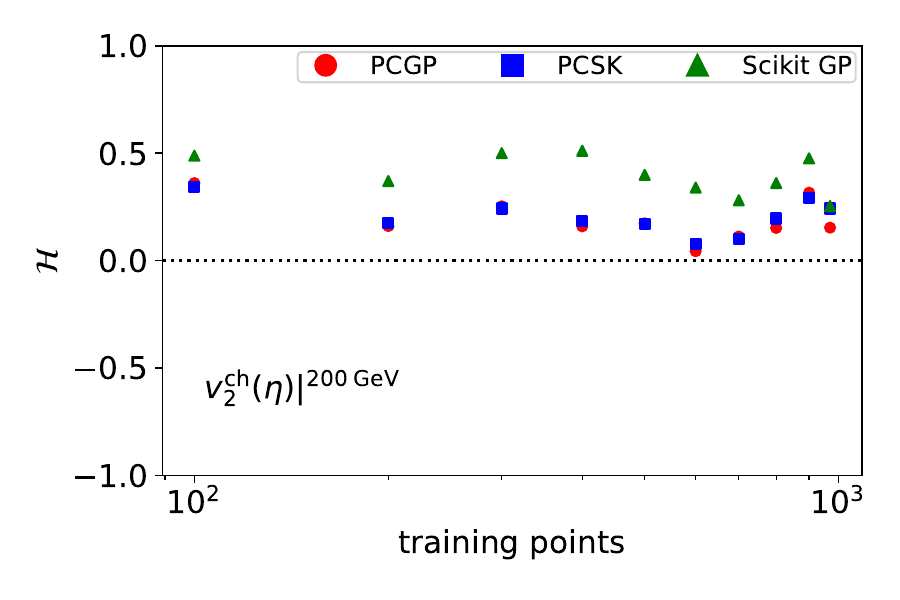}
    \caption{(Color Online) The averaged RMS errors for emulators $\mathcal{E}$ (left) and the quality of uncertainty estimation measure $\mathcal{H}$ (right) for three different example observables. The top row shows the first observable from the 7.7 GeV $\langle p_{\rm T}\rangle$ dataset, and the middle row is taken from the first observable in the 200 GeV ${\rm d}N/{\rm d}y$ dataset. The bottom row shows an observable at midrapidity from the 200 GeV $v_2(\eta)$ dataset.}
    \label{fig:examples_nTrain_dependence}
\end{figure*}

Figure~\ref{fig:examples_nTrain_dependence} shows for three different observables at two different energies the emulator RMS error $\mathcal{E}$ (left) defined in Eq.~\eqref{eq:uncertainty} and the quality measure of the uncertainty estimation $\mathcal{H}$ (right) defined in Eq.~\eqref{eq:honesty}.
In the first row, we show the first observable from the 7.7 GeV $\langle p_{\rm T}\rangle$ dataset; in the second one, the first observable from the 200 GeV ${\rm d}N/{\rm d}y$ dataset, and in the last one the $v_2(\eta)$ at midrapidity from the 200 GeV dataset.
We divide the 1,000 design points into two groups: the first $n_{\rm train}$ points for training the GP emulators and the rest for validation.
Because we use the Maximum Projection Latin Hypercube Design, the first $n_{\rm train}$ points can give a uniform coverage of the 20-dimensional model parameter space.
We note that as $n_{\rm train}$ increases to close to the full dataset, the limited number of validation points would introduce statistical fluctuations when computing the GP test metrics $\mathcal{E}$ and $\mathcal{H}$.
Therefore, we would expect good scaling behavior for $n_{\rm train}$ up to 700 in our test.

Starting with the $\langle p_{\rm T}\rangle$ observable (top panels), we can see that increasing the training data size decreases the emulation uncertainty with a power-law scaling behavior, inline with previous results~\cite{Nijs:2020ors}. 
The quantitative power law dependence varies from one observable to the other, therefore we do not report any fit values here.
The value for the quality measure of the GP uncertainty estimation $\mathcal{H}$ decreases from 0.5 to $\approx 0$ for $n_{\rm train} \ge 300$.
It shows that the GP emulators can provide a good estimation of their prediction uncertainty.
We can also see that the PCGP and PCSK emulators give a smaller uncertainty overall and better uncertainty estimation compared to the Scikit GP emulator.

The test results for ${\rm d}N/{\rm d}y$ (middle panels) show an even more clear separation between the Scikit GP emulator and the other two emulators, with almost a factor of two larger uncertainty for $n_{\rm train} \ge 700$.
The $\mathcal{H}$ metric is close to zero, but the PCGP and PCSK emulators even give a more conservative estimate while having a smaller prediction uncertainty.

The $v_2(\eta)$ observable (bottom panels) shows a similar performance of all three emulators in the uncertainty estimate. Still, the PCGP and PCSK emulators seem to be slightly better at estimating the prediction uncertainty compared to the Scikit GP emulator.

\subsubsection{Benchmark Emulation Performance}
\label{sub:emu_uncertainties}

\begin{figure*}[tb!]
    \centering
    \includegraphics[width=\textwidth]{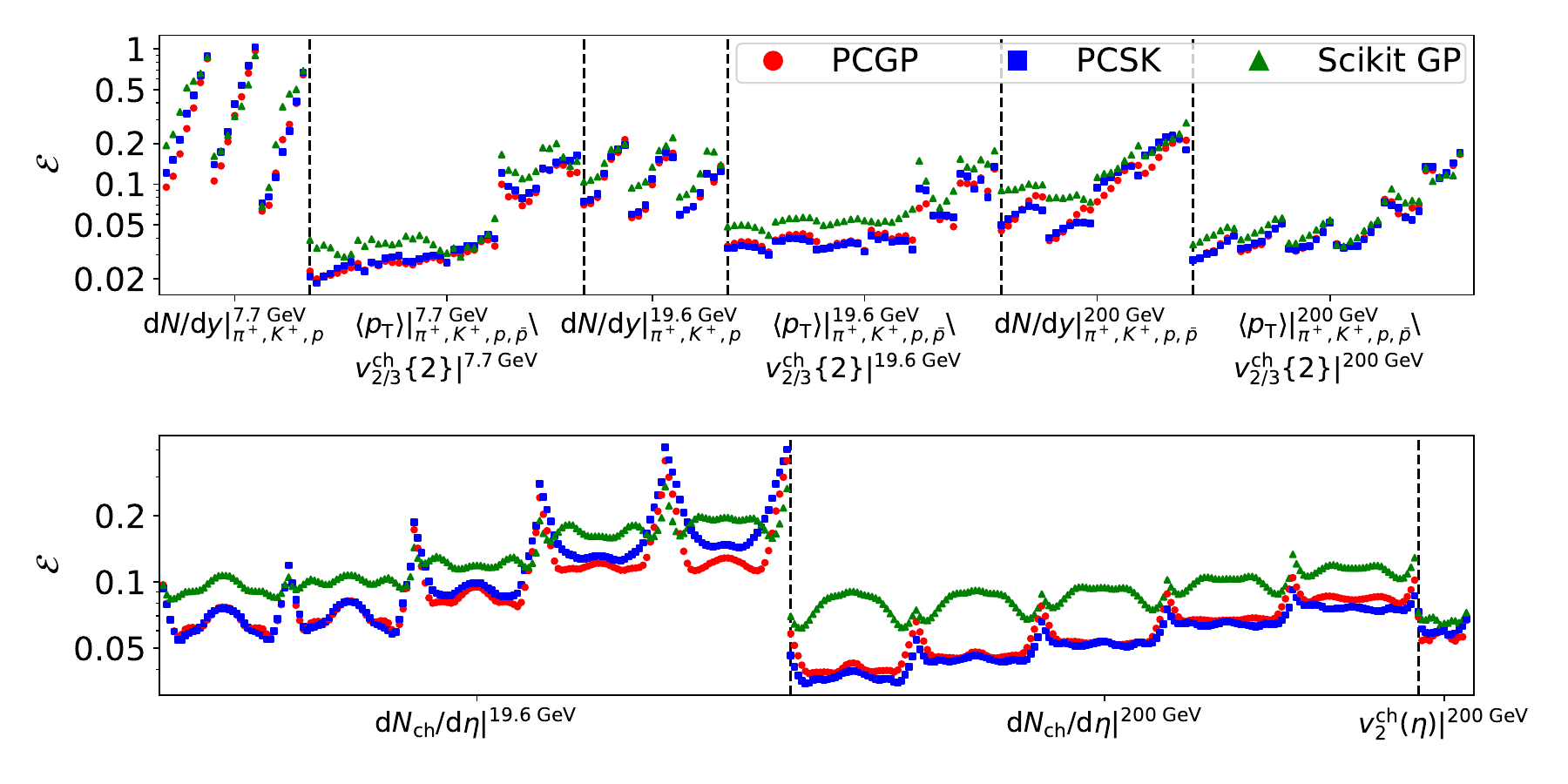}
    \caption{(Color Online) The averaged RMS errors $\mathcal{E}$ for the three different types of GP emulators. Different training sets are separated by black lines. All emulators are trained with the same 970 LHD points.}
    \label{fig:E_GP}
\end{figure*}
\begin{figure*}[tb!]
    \centering
    \includegraphics[width=\textwidth]{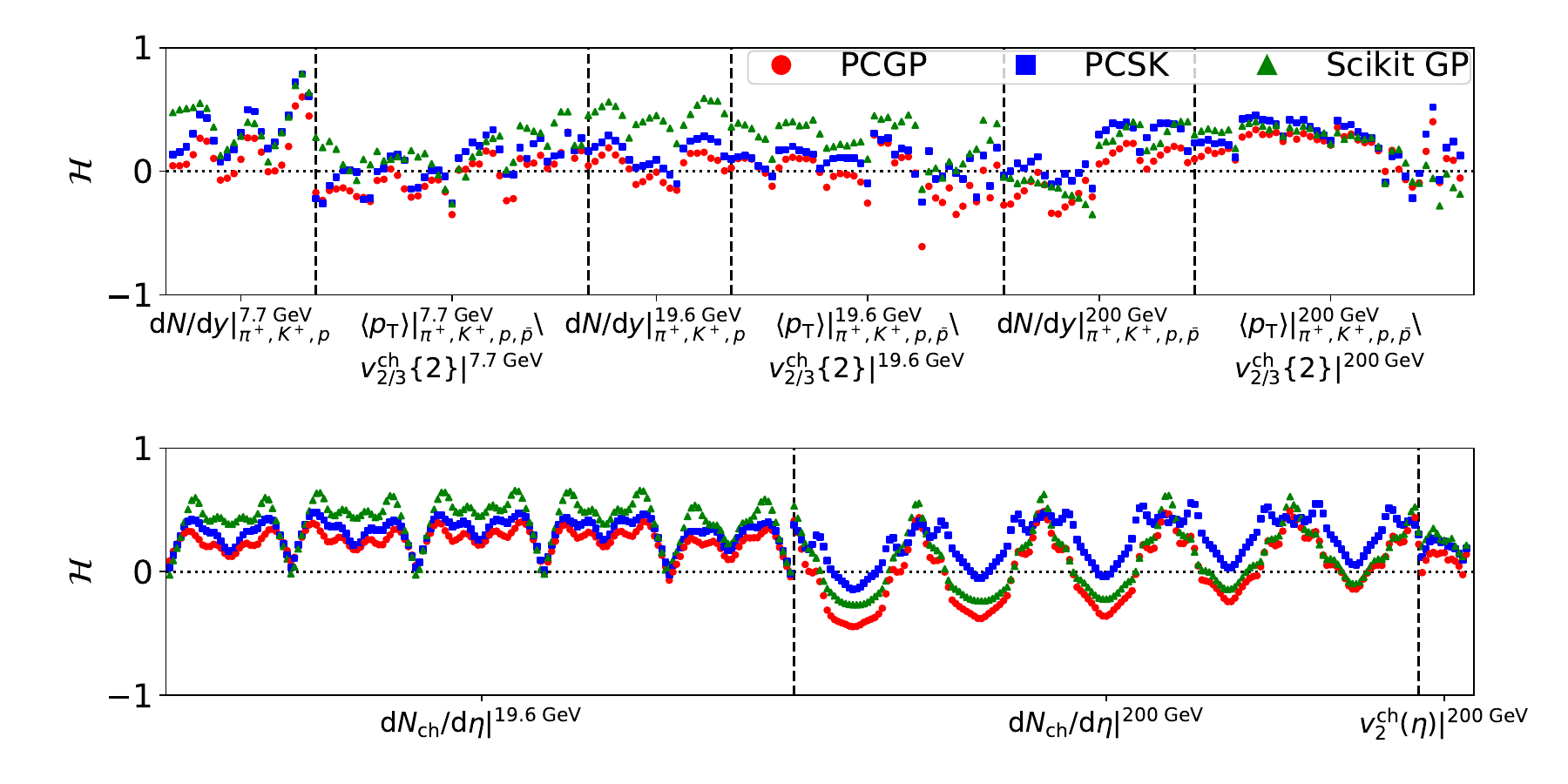}
    \caption{(Color Online) The metric for emulator uncertainty estimation $\mathcal{H}$ for the three different types of GP emulators. Black lines separate different training sets. All emulators are trained with the same 970 LHD points.}
    \label{fig:H_GP}
\end{figure*}

To benchmark and compare the performance of the three types of GP emulators over all observables, we now fix the number of training points from the LHD to 970 in the following.
This leaves us 30 validation points to compute the $\mathcal{E}$ and $\mathcal{H}$ test metrics for the emulators. 
The absolute values of $\mathcal{E}$ and $\mathcal{H}$ will still contain some statistical fluctuations.

Figure~\ref{fig:E_GP} displays the emulators' RMS error $\mathcal{E}$ for all 544 experimental observables -- separated by black lines according to Table~\ref{tab:training_data} -- trained with the three types of GP emulators.
The upper panel summarizes the mid-rapidity observables from central to 60\% peripheral centrality for Au+Au collisions at three energies.
The lower panel shows the pseudo-rapidity dependence observables in five centrality classes at 19.6 and 200 GeV.
For the ${\rm d}N_{\rm ch}/{\rm d}\eta$ datasets, we can identify the different ``$w$''-shaped structures, which are the multiplicities as a function of rapidity in different centrality classes.
Each point in the plot represents a single observable, indicated by the ticks on the $x$-axis.

We observe that the PCGP and PCSK emulators are compatible with each other for most observables and that their overall prediction uncertainty is lower than that of the Scikit GP.
This trend was already visible in comparing the emulators for single observables as a function of the number of training points in Sec.~\ref{subsub:train_point_dependence}.
The particle yield $\mathrm{d}N/\mathrm{d}y$ dataset at $7.7\;\mathrm{GeV}$ seems to be equally challenging for all three emulators, with uncertainties up to 100\% in peripheral centrality bins, while most other points are in the region of $2-20\%$.

Next, we want to analyze how reliable the uncertainty provided by the emulators is with the metric $\mathcal{H}$ defined in Eq.~\eqref{eq:honesty}.
The results for the three types of GP emulators are shown in Fig.~\ref{fig:H_GP}, using the same conventions as in Fig.~\ref{fig:E_GP}.
When $\mathcal{H}$ is around zero, the estimated uncertainty is close to the true uncertainty in our validation test.

For all three types of GP emulators, most observables show a small trend toward positive values, which means that the emulator underestimates the uncertainty compared to the true values in the validation set.
Comparing the GP emulators, we find that the PCGP and PCSK emulators perform better than the Scikit GP emulator.
Their $\mathcal{H}$ are closer to 0 than those from the Scikit GP emulator.

Overall, we find the PCGP and PCSK emulators from the BAND \texttt{surmise} package perform better than the Scikit GP emulator with standard settings.
We note that the relatively poorer performance of the Scikit GP emulator could be related to not having optimized the GP kernel and its parameters.\footnote{The Scikit GP kernel consists of an RBF kernel with an additional white noise term similar to Eq.~(4.29) in~\cite{Bernhard:2018hnz}. The choice of default kernel parameters followed the JETSCAPE calibration~\cite{JETSCAPE:2020mzn}.}

\subsubsection{Logarithmic transformation to training datasets}
\label{subsub:log_training}

\begin{figure*}[tb!]
    \centering
    \includegraphics[width=\textwidth]{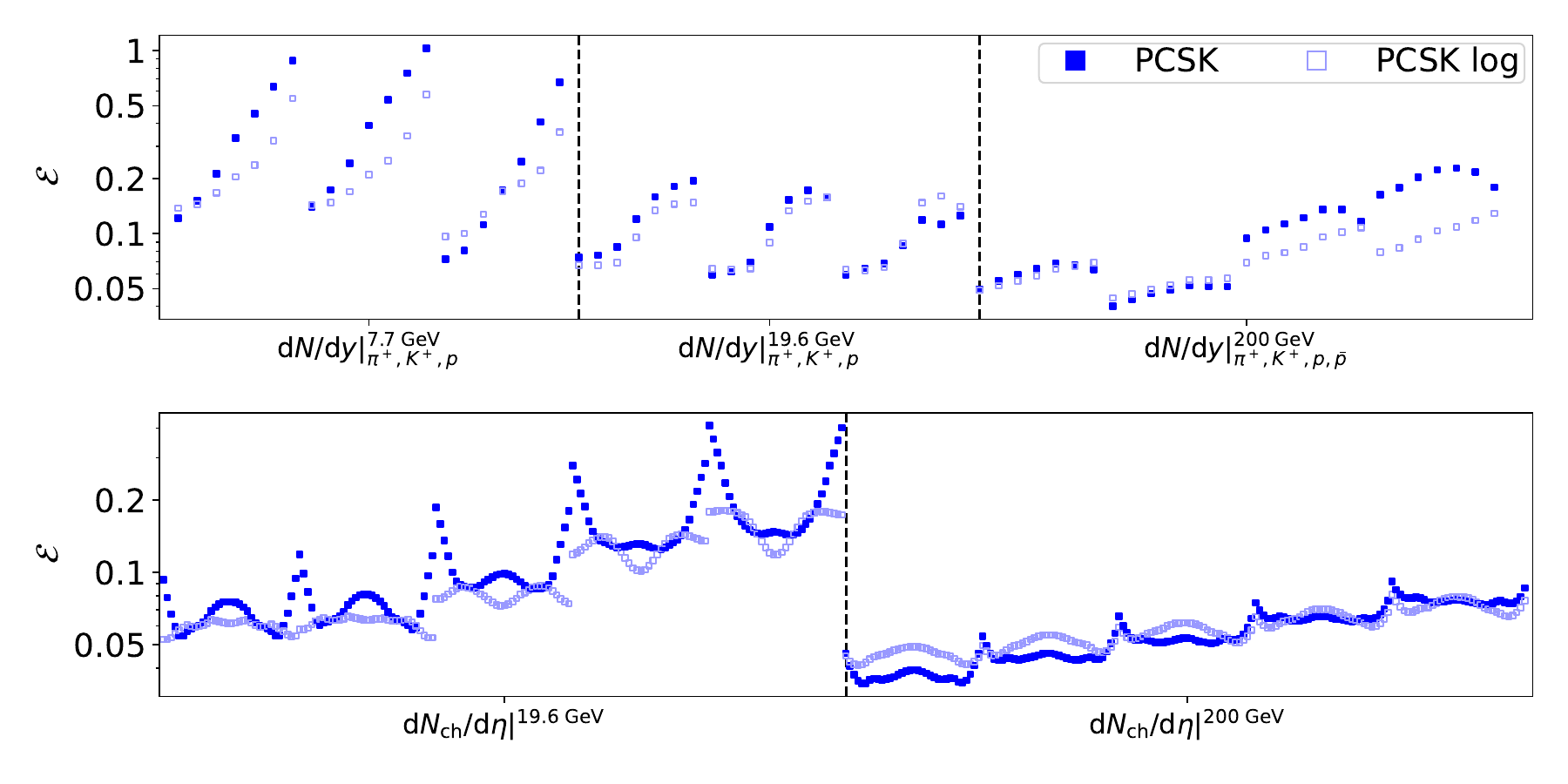}
    \caption{(Color Online) The averaged RMS error $\mathcal{E}$ for the PCSK emulator trained with (open symbols) and without (solid markers) logarithmic transformation to particle multiplicities observables. Both emulators are trained with the same 970 LHD points and validated with the remaining 30 points in the initial design. Black lines separate different sets of observables.}
    \label{fig:E_GP_LOG_PCSK}
\end{figure*}
\begin{figure*}[tb!]
    \centering
    \includegraphics[width=\textwidth]{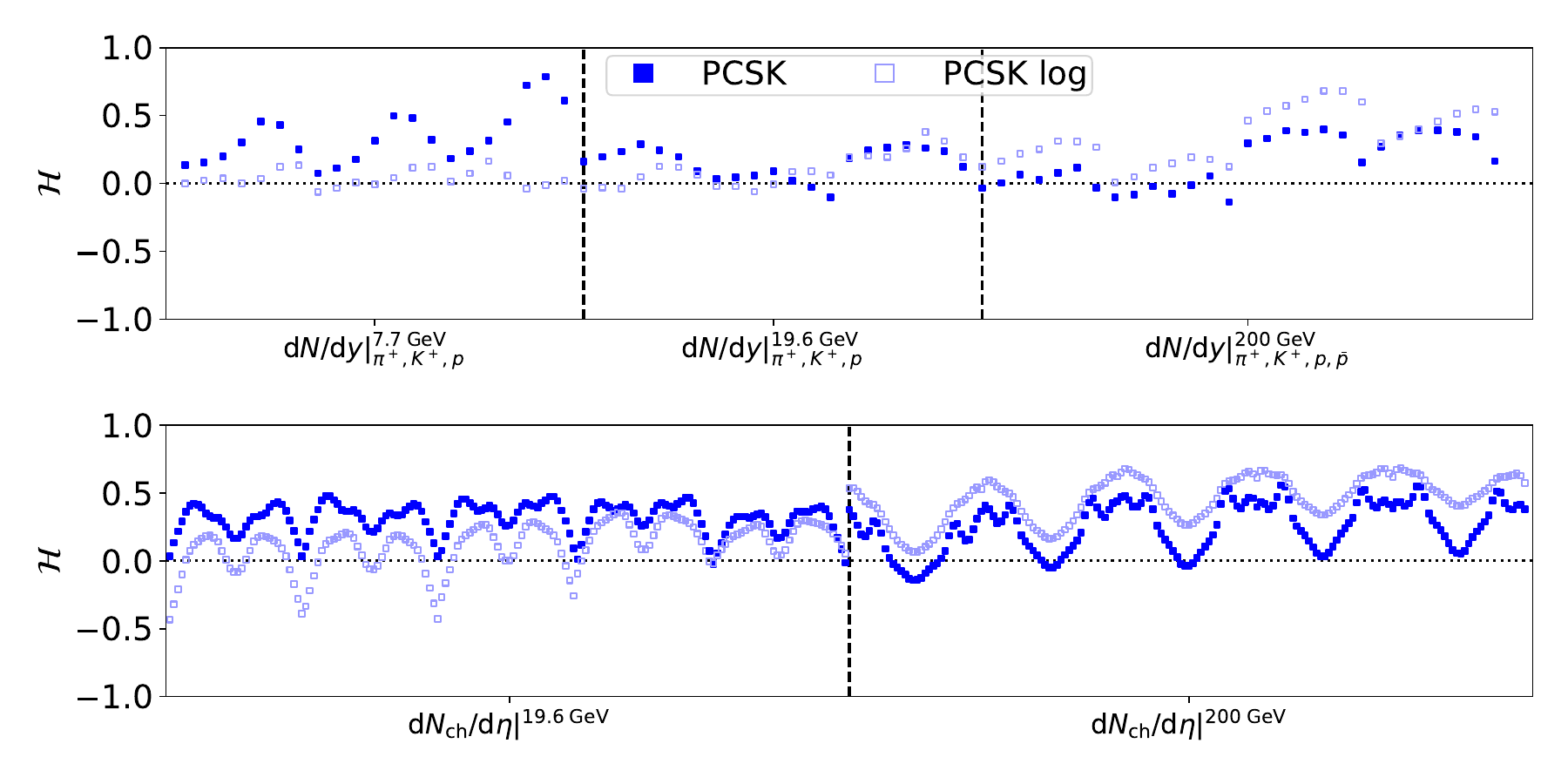}
    \caption{(Color Online) Same layout as in Fig.~\ref{fig:E_GP_LOG_PCSK} but for the metric for emulator uncertainty estimation $\mathcal{H}$.}
    \label{fig:H_GP_LOG_PCSK}
\end{figure*}

Based on the results from previous sections, we now pick the PCSK emulator for further investigation because it is one of the two more accurate ones and incorporates the uncertainties of the training data.
In Appendix~\ref{app:PCSK_err}, we discuss the emulation results of this emulator in more detail concerning the size of the uncertainties of the training points and how it compares to the PCGP emulator in the case no uncertainties are provided.
The results discussed in the following are general and hold for the other two emulators as well (not shown).

In this section, we discuss the impacts of training with the logarithm of the multiplicity observables on the accuracy of the GP emulations.
Performing logarithmic transformation on particle multiplicities was found to be an effective trick for yielding more accurate GP emulators~\cite{Nijs:2020ors, JETSCAPE:2020mzn}.
It is suspected that $\ln(\rm{d}N/\rm{d}y)$ reduces the relative variations in the training dataset. 
However, the scaling transformation described in Sec.~\ref{sec:GP} should be a more systematic way to adjust the relative variations in the training data.

Figure~\ref{fig:E_GP_LOG_PCSK} shows in full squares the $\mathcal{E}$ points for PCSK trained directly with particle multiplicities $\rm{d}N/\rm{d}y$ (from Fig.~\ref{fig:E_GP}) and compares them to the results from the logarithmic trained emulators shown in open light blue points.

We find that the uncertainty is decreased for almost all five datasets or remains comparable to the standard results.
Only for very few points does the uncertainty increase by a few percent.
The largest decrease is visible for the large rapidity data points, which are the rightmost ones in the linear rise of the $\mathrm{d}N/\mathrm{d}y$ datasets and the outermost points in the ``w''-shaped parts of the $\mathrm{d}N_{\rm ch}/\mathrm{d}\eta$ datasets.

Concerning the metric $\mathcal{H}$ for the uncertainty estimate in Fig.~\ref{fig:H_GP_LOG_PCSK}, we find that for most datasets the points move closer to zero or stay reasonably close to zero.

Overall, we find that the logarithmic transformation to the training data can be considered a valid way to decrease the uncertainty of quantities varying over a large range.

\subsubsection{PCA Transformation to Functional Model Parameters}
\label{subsub:PCA_trafo}

\begin{figure*}[tb!]
    \centering
    \includegraphics[width=\textwidth]{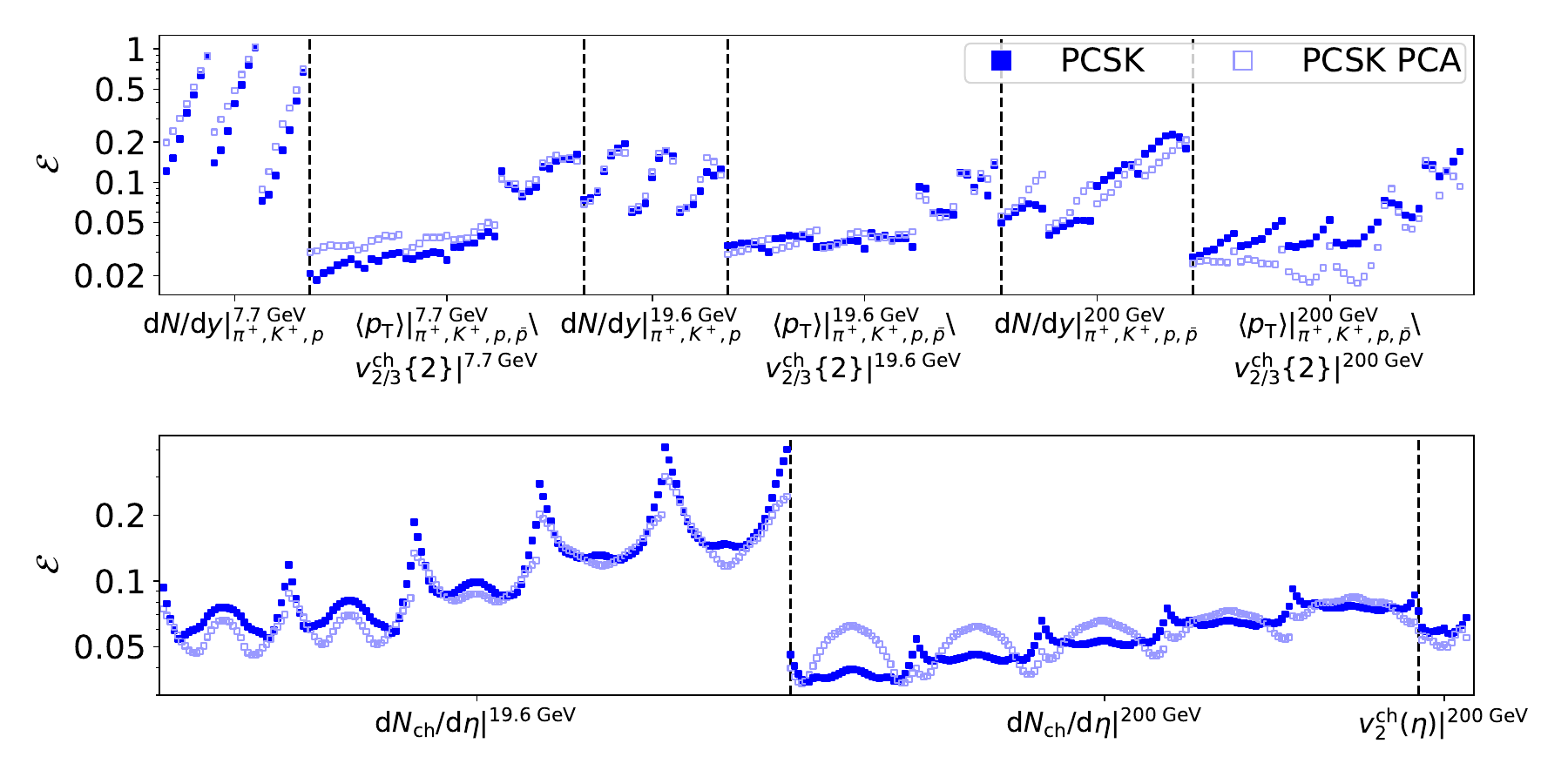}
    \caption{(Color Online) The same layout as in Fig.~\ref{fig:E_GP} but for comparing the PCSK emulators with and without PCA transformation on the functional model parameters.}
    \label{fig:E_GP_PCA_PCSK}
\end{figure*}
\begin{figure*}[tb!]
    \centering
    \includegraphics[width=\textwidth]{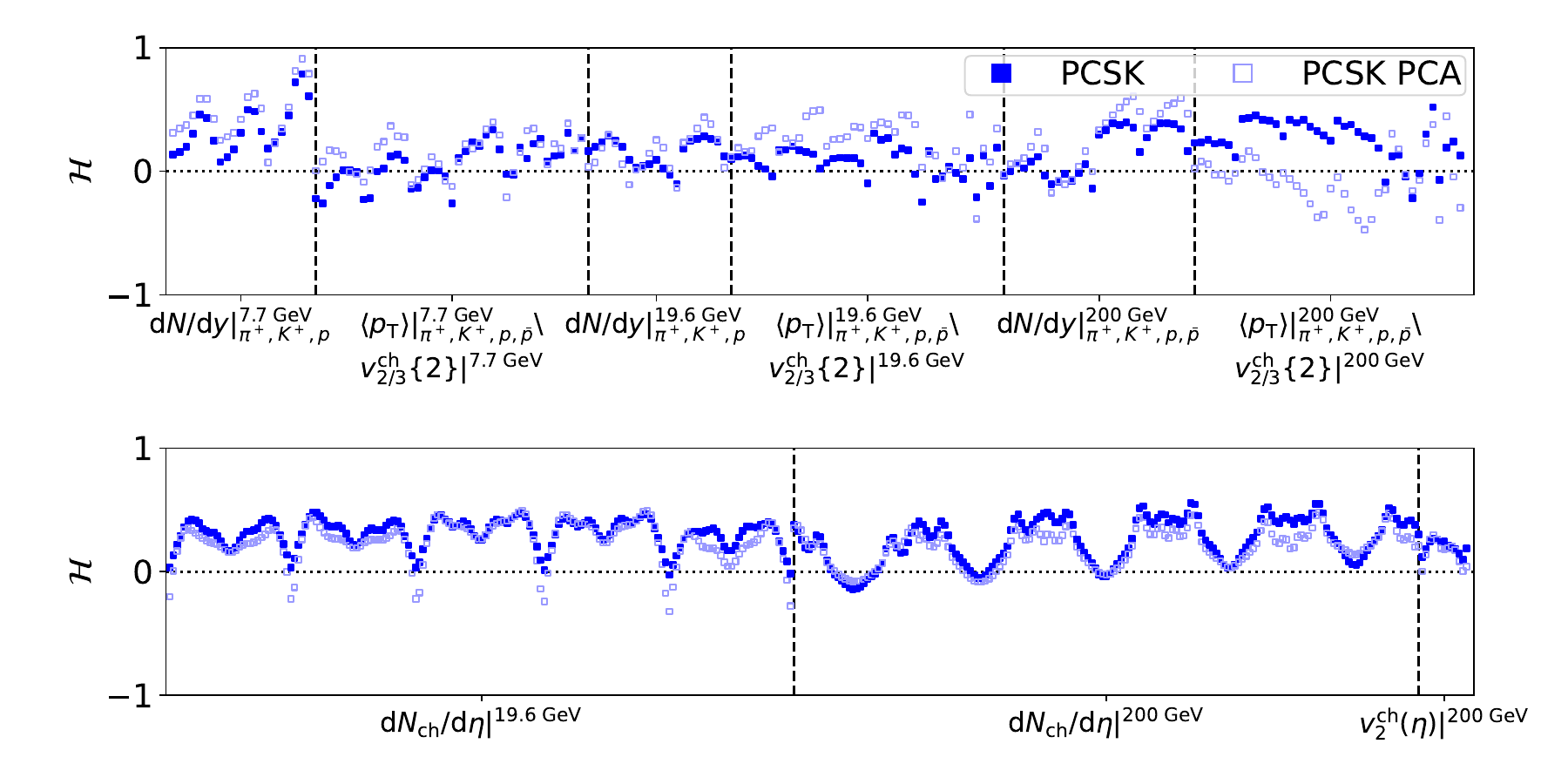}
    \caption{(Color Online) The same layout as in Fig.~\ref{fig:E_GP_PCA_PCSK} but for the metric for emulator uncertainty estimation $\mathcal{H}$.}
    \label{fig:H_GP_PCA_PCSK}
\end{figure*}

In this section, we will assess the impacts on the emulator accuracy from performing PCA transformations on the functional model inputs, $\tilde{\zeta}(T,\mu_B)$, $\tilde{\eta}(\mu_B)$ and $\langle y_{\rm loss}\rangle(y_\mathrm{init})$, and use the principal components to fit the GP emulators instead of the curve parameters listed in Table~\ref{tab:parameters}.

To explain 99\% of the variance in the training design, we need three principal components for $\tilde{\eta}(\mu_B)$ and $\langle y_{\rm loss}\rangle(y_\mathrm{init})$ and six principal components for $\tilde{\zeta}(T,\mu_B)$.
Since the functional parametrizations for $\tilde{\eta}(\mu_B)$ and $\langle y_{\rm loss}\rangle(y_\mathrm{init})$ are simple piece-wise linear functions (see Eqs.~\eqref{eq:ylossParam} and \eqref{eq:QGPshearParam}), the number of principal components needed is the same as the number of model parameters.
For the QGP's specific bulk viscosity, the parameterization in Eq.~\eqref{eq:QGPbulkParam} is non-linear.
The PCA transformation indicates that two more degrees of freedom are needed in addition to the four model parameters.
We checked that the averaged relative RMS errors are less than 5\% between the original parameterizations and the ones reconstructed using the truncated principal components.

Figure~\ref{fig:E_GP_PCA_PCSK} shows the standard PCSK emulators in full points and the results (in open markers) from the PCSK GP trained with the principal components from the PCA transformations on the model parameters, $\tilde{\zeta}(T,\mu_B)$, $\tilde{\eta}(\mu_B)$ and $\langle y_{\rm loss}\rangle(y_\mathrm{init})$.
We note that because of the PCA transformation on model parameters, our training points may no longer be evenly distributed in the 22-dimensional model parameter space.

Comparing the full and open points across the whole set of observables, we find that the PCA transformations do not significantly alter the overall emulation performance.
There are a few observables where the uncertainty increases by a few percent, but there are also observables with a decreased uncertainty. 

The same statement holds for the metric for emulator uncertainty estimation shown in Fig.~\ref{fig:H_GP_PCA_PCSK}, where overall the $\mathcal{H}$ values are not significantly affected by the PCA transformation.

Our results indicate that the relevant model parameters listed in Table~\ref{tab:parameters} are sufficient to capture all the essential information coded in Eqs~\eqref{eq:ylossParam}, \eqref{eq:QGPshearParam}, and \eqref{eq:QGPbulkParam}, when training the GP emulators.
The two additional degrees of freedom in the PCA transformed model parameter space do not significantly influence the GP performance.

However, it is noteworthy, that the PCA transformation we introduced here could serve as a generic approach towards feeding general functional model inputs to the GP emulators.

In the future, this approach can be used to allow for a parameter-less emulation of the rapidity loss and the viscosities in the hydrodynamic evolution.

\subsubsection{Active Learning Approach}
\label{subsub:active_learning}

\begin{figure*}[tb!]
    \centering
    \includegraphics[width=\textwidth]{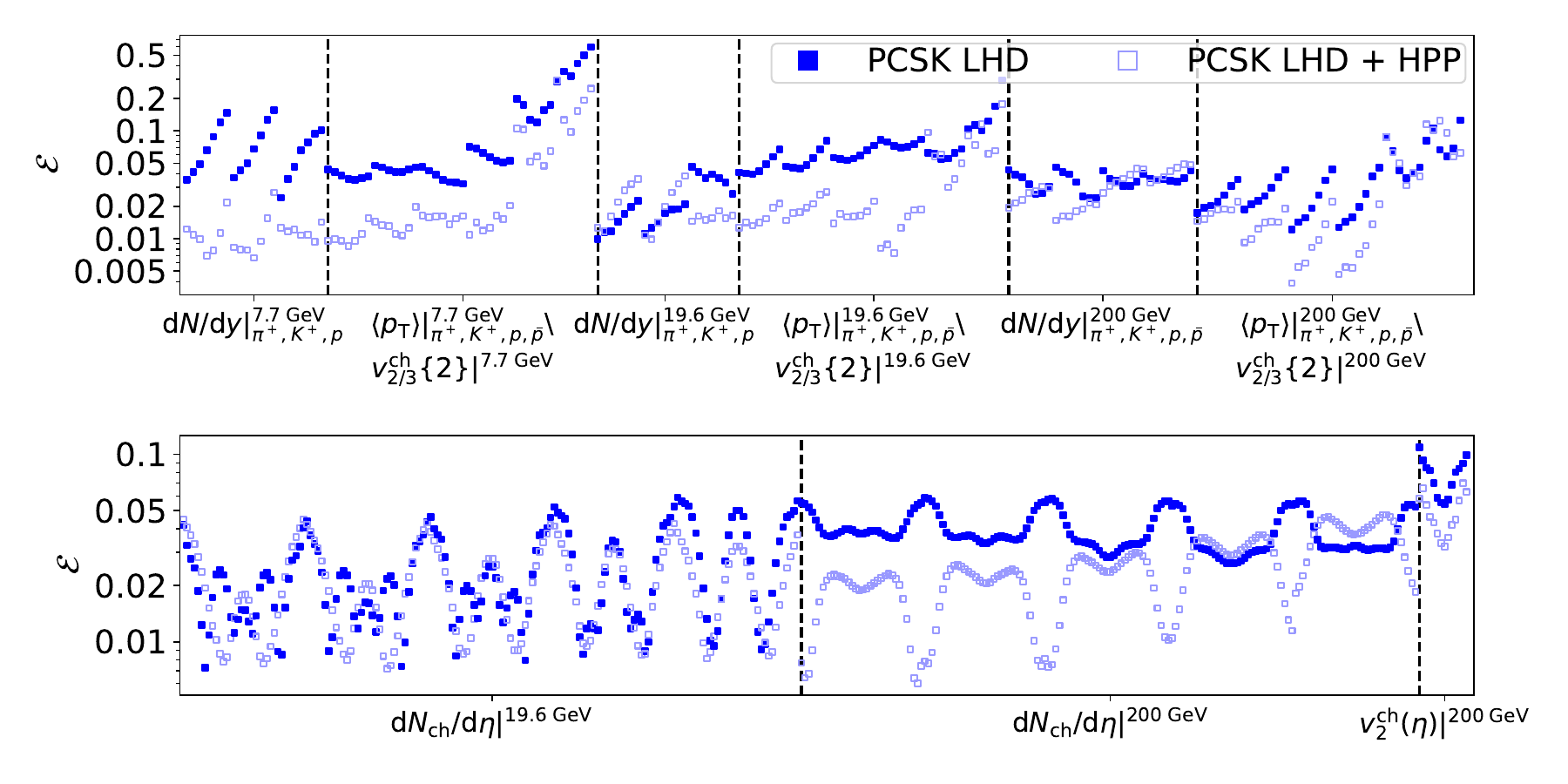}
    \caption{(Color Online) The same layout as in Fig.~\ref{fig:E_GP} but for comparing the PCSK emulators trained with and without the high probability posterior points.}
    \label{fig:E_GP_active_learning}
\end{figure*}
\begin{figure*}[tb!]
    \centering
    \includegraphics[width=\textwidth]{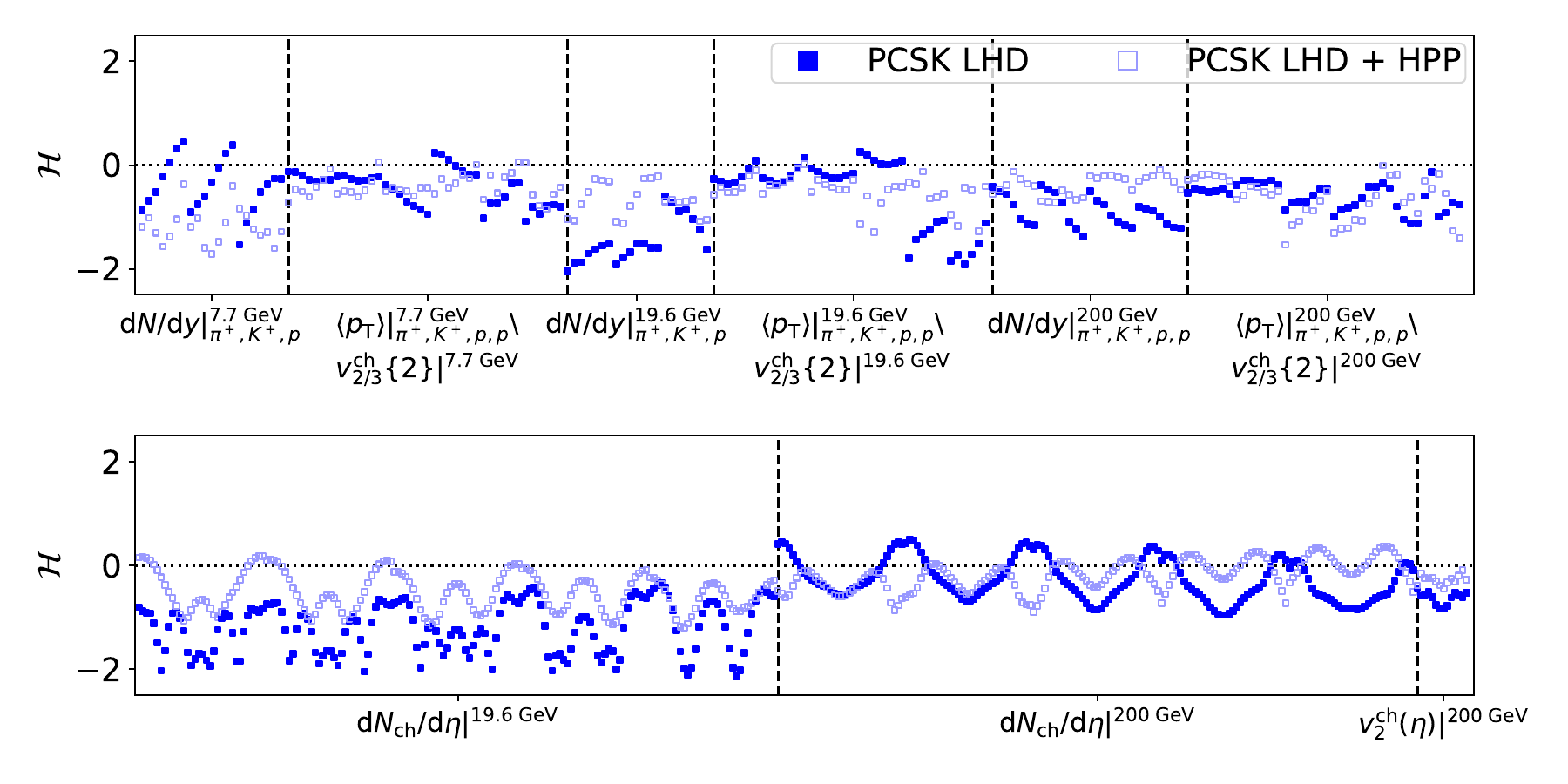}
    \caption{(Color Online) The same layout as in Fig.~\ref{fig:E_GP_active_learning} but for the metric for emulator uncertainty estimation $\mathcal{H}$.}
    \label{fig:H_GP_active_learning}
\end{figure*}

The dimension of our model parameters is high.
It is more computationally efficient to reduce the emulator uncertainty only in the region of model space near where the real experimental data lie than across the entire parameter space. 

For this purpose, we explore the following two-step approach for our emulator training.
First, we train a model emulator with the given 1000 LHD points and obtain the model's posterior distribution by performing a Bayesian inference analysis on the real experimental data~\cite{Shen:2023awv}.
Then we run full model simulations on 100 parameter sets sampled from the posterior distribution.
We denote this dataset as the High Probability Posterior (HPP) points in the following.
Finally, we train our GP emulators using the combined LHD + HPP datasets.
Because the HPP points populate a relatively dense region near the physical posterior region, they would improve the emulator's accuracy within this phase space.

In the following tests, we train the GP emulators using 1000 LHD + 70 HPP points and evaluate their performance in the model posterior region with the rest 30 HPP points as the validation data.
We compare such a setup with the standard GP training with only the 1000 LHD points.

In this section, we only evaluate the emulator performance by computing the $\mathcal{E}$ and $\mathcal{H}$ metrics on the leftover 30 HPP points.
The values of the $\mathcal{E}$ and $\mathcal{H}$ are different from the results from previous sections, in which the validations were performed over the entire model parameter space.

We denote our approach here as ``active learning'' although we only have one iteration in our exercise.
In principle, this technique can be applied in multiple iterations to increase the accuracy in the interesting regions of the high dimensional parameter space.
This allows reducing the computational cost in a Bayesian parameter extraction using a computationally expensive physics model.
A similar approach was employed in previous works~\cite{JETSCAPE:2023ikg, Liyanage:2023nds}. 

Figure~\ref{fig:E_GP_active_learning} shows the comparison between the fully trained LHD data set (full symbols) and the active learning (open symbols). 

First, we can say that the overall uncertainty of all quantities decreases significantly when the full 1000 LHD points are taken into account and the emulator is tested with the posterior points.
This is probably due to a bias in the testing in only a preferred region of the parameter space, where the posterior is peaked. 
Nevertheless, we can see that the inclusion of the posterior points improves almost all observables.
The ones that do not change have similar uncertainty compared to the standard setup, while others change by up to factors five to ten.

Figure~\ref{fig:H_GP_active_learning} shows that using only the posterior points to compute $\mathcal{H}$ leads to more conservative uncertainty estimates of the emulator ($\mathcal{H}<0$), which is of similar order for most of the observables and both methods.

Let us also mention here that including too many HPP points when training GP emulators can significantly reduce the emulator uncertainty in certain phase space regions. The non-uniform size of the emulator uncertainty over the entire parameter space could potentially introduce bias (small likelihood) to the parameter space near HPP when performing Bayesian inference with real experimental measurements~\cite{Jahan:2024wpj}.
Based on the improvement in our closure test, we are confident that adding a small amount of HPP points ($\sim 10\%$ of the LHD points used) for the training does not introduce a noticeable bias in the Bayesian inference.

\section{Closure Tests}
\label{sec:MCMC}

In this section, we will explore how the different levels of emulator performance can influence the model parameter posterior distribution in the Bayesian inference analysis.

\subsection{MCMC Setups}
\label{sub:mcmc_setup}
Our analysis in this paper utilizes two different implementations of posterior sampling techniques.
The first employs affine invariant MCMC, leveraging the \texttt{emcee} Python package~\cite{Foreman_Mackey_2013}.
Secondly, we employ Parallel Tempering Langevin Monte Carlo (PTLMC), implemented in the \texttt{surmise} package~\cite{surmise2023}.
The parallel tempering part enhances the exploration of the target distribution by running multiple chains at different ``temperatures'' and exchanging states periodically, which scans the parameter space more accurately and reduces the probability of getting stuck in a local minimum.
This technique is of advantage if the posterior distribution is a multi-modal distribution and additionally converges faster than without parallel tempering.
Additionally, some Langevin dynamics are incorporated to explore complex posterior distributions more efficiently~\cite{Roberts2002-om}.

To test the different scenarios, including the different emulators and methods discussed in Sec.~\ref{sub:emu_validation}, we are defining some Runs in Tab.~\ref{tab:mcmc_setup}.
The first three setups in our analysis compare the impact of different types of GP emulators on posterior distributions in a standard configuration.
This configuration excludes logarithmic transformation, but incorporates PCA for the functional model inputs, and trains the emulators using both LHD and HPP points.
By examining these Runs, we aim to assess how various emulator choices influence the accuracy and reliability of the posterior inference process.

Run 4 mirrors the Run 1's setup but utilizing PTLMC to obtain the posterior.
This direct comparison between MCMC and the parallel tempering variant allows us to evaluate the efficacy of temperature-enhanced exploration techniques in improving posterior sampling efficiency and exploring high-dimensional parameter spaces more effectively.
By employing the advanced sampling method, we seek to uncover insights into how different sampling strategies impact the exploration of posterior distributions and the convergence properties of Bayesian inference algorithms.

By contrasting Runs 1 with 5, we can elucidate the influence of incorporating an additional 95 HPP points from the posterior distribution in the training phase on posterior inference.
Similarly, comparing Runs 1 with 6 and 7 enables us to quantify the impact of PCA and logarithmic transformation in the training design on the posterior distributions.

These comparative analyses offer valuable insights into the various factors that contribute to the accuracy and reliability of the GP emulators that affect Bayesian posterior sampling techniques.

\subsection{Closure Test}
\label{sub:closure_test}

In Section~\ref{sub:emu_validation}, we demonstrated that the emulators can predict the training data with a quantifiable level of uncertainty, as discussed previously.
However, the ability of the given data sets to effectively constrain model parameters may vary.
If the data exhibit insensitivity to certain parameters, their constraints may be poor, leading to broad posterior distributions.
Additionally, if the underlying physics model contains degeneracies, where different parameter sets can equally describe the experimental data, posterior distributions may even become multimodal.
To identify such scenarios, conducting a closure test is needed.
In a closure test, one training point is excluded from the training set and treated as the pseudo-experimental data.
By carrying out Bayesian inference analysis on the pseudo-data, we can compare the obtained posterior distributions with the true model parameters used to generate the pseudo-data.
Ideally, the posterior distributions should exhibit peaks close to the true value.
In cases of model degeneracy, the peaks should at least be proximate.
By doing closure tests, we can quantify how well our method works in understanding the model and the limitations of our particular Bayesian inverse problem.

\begin{figure*}[tb!]
    \centering
    \includegraphics[width=\textwidth]{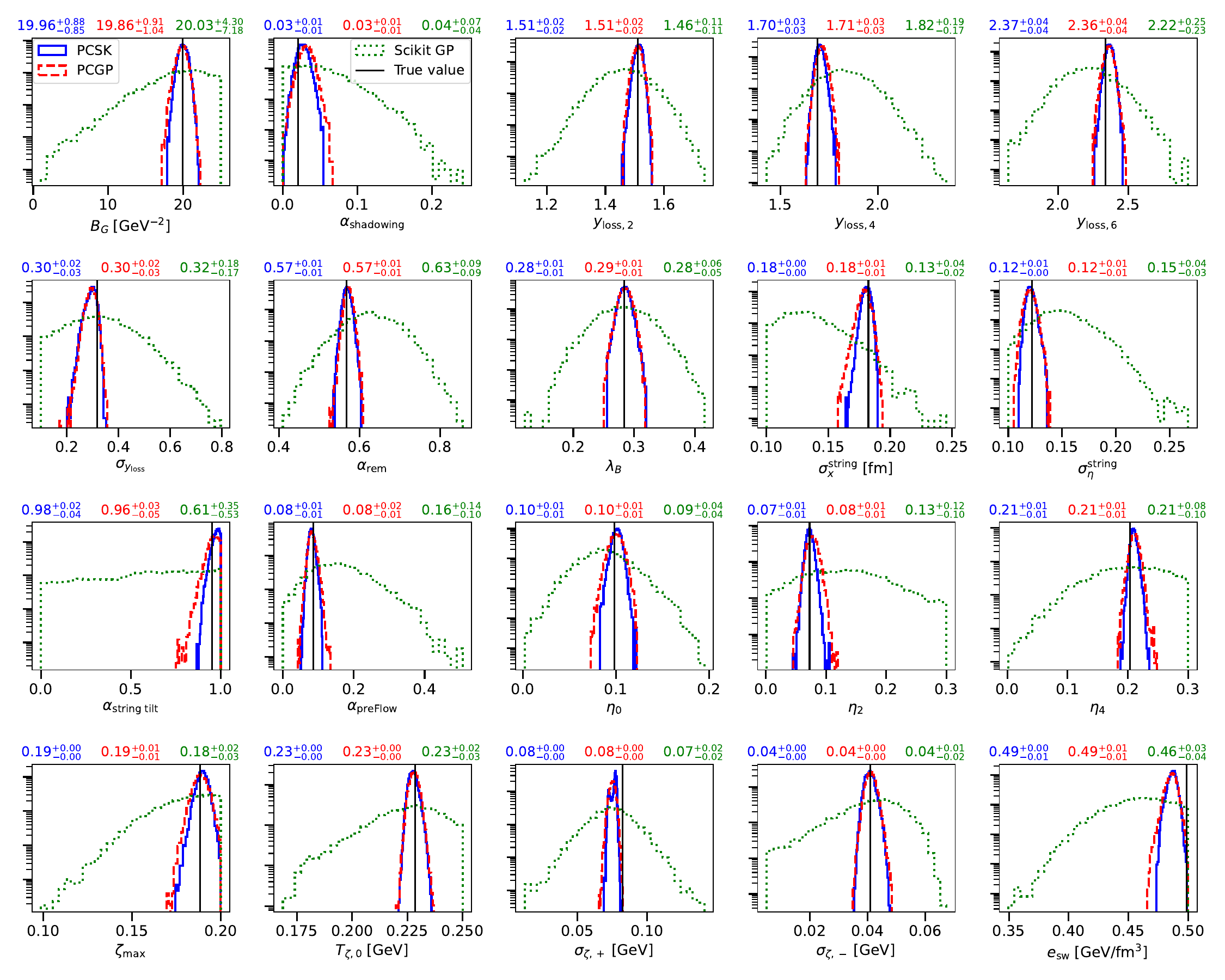}
    \caption{(Color Online) Closure test posterior distributions for the 20 different parameters from Tab.~\ref{tab:parameters} using the three different GP emulators with runs 1, 2, and 3 in Tab.~\ref{tab:mcmc_setup}. The marginalized parameter distributions are plotted in logarithmic scale in the $y$-axis. The true value is indicated by the solid black line. The values above each plot in the respective colors of the lines are the median values with the 90\% confidence level intervals.}
    \label{fig:posterior_R123}
\end{figure*}

Figure~\ref{fig:posterior_R123} shows the comparison of the three different types of GP emulators from Runs 1, 2, and 3 listed in Tab.~\ref{tab:mcmc_setup}.
We find that the PCSK and PCGP emulators show clear and narrow peaks around the true values indicated by the black vertical line and relatively a lot broader posterior distributions for the results using the Scikit GP emulator.
This result shows that the quality of the GP emulator has a strong influence on the posterior distribution extracted in the Bayesian inference.
The PCSK and PCGP emulators show compatible results in the closure test.

To further quantify the closure test and compare results from different setups, introduced in Tab.~\ref{tab:mcmc_setup} we will compare a metric $\Delta$ from the information-theoretic approach similar to Ref.~\cite{JETSCAPE:2023ikg}.
The main contribution to the GP emulator information loss, which is mainly determined by the second moment of the posterior distribution, using
\begin{equation}
    \Delta \equiv \frac{1}{N_{\rm param.}}\int \left|\frac{\bs{\theta}-\bs{\theta}_{\rm truth}}{\bs{\theta}_{\rm max}-\bs{\theta}_{\rm min}}\right|^2 p(\bs{\theta})\;\mathrm{d}\bs{\theta},
    \label{eq:delta_d}
\end{equation}
where $\bs{\theta}_{\rm truth}$ is the closure test point, $|\bs{\theta}_{\rm max}-\bs{\theta}_{\rm min}|$ is the allowed prior range and $p(\bs{\theta})$ is the posterior distribution, where the closure test point is not included.
Different from the definition in Ref.~\cite{JETSCAPE:2023ikg}, we are dividing by $N_{\rm param.}$, the number of model parameters, to normalize the values of $\Delta$ between zero and one.
The smaller the value, the more peaked is the posterior distribution around the true value.
The last column of Tab.~\ref{tab:mcmc_setup} summarizes the values of $\Delta$ for the different MCMC runs.
\begin{table*}[tb!]
    \caption{Closure test runs with GP emulators trained in different settings. The metric $\Delta$ in the last column quantifies the quality of the obtained posterior distributions to the true model values in the closure test.}
    \label{tab:mcmc_setup}
    \centering
    \begin{tabular}{c|c|c|c|c|c|c}
        \hline\hline
        Run & Training Design & MCMC Type & Log Transformation & PCA Transformation & Emulator & $\Delta$ \\
        \hline
        1 & LHD + HPP & MCMC & $\times$ & $\checkmark$ & PCSK & $4.5\times 10^{-4}$ \\
        2 & LHD + HPP & MCMC & $\times$ & $\checkmark$ & PCGP & $5.7\times 10^{-4}$ \\
        3 & LHD + HPP & MCMC & $\times$ & $\checkmark$ & Scikit GP & $2.5\times 10^{-2}$ \\
        4 & LHD + HPP & PTLMC & $\times$ & $\checkmark$ & PCSK & $4.4\times 10^{-4}$ \\
        5 & LHD & MCMC & $\times$ & $\checkmark$ & PCSK & $4.5\times 10^{-2}$ \\
        6 & LHD + HPP & MCMC & $\times$ & $\times$ & PCSK & $4.3\times 10^{-4}$ \\
        7 & LHD + HPP & MCMC & $\checkmark$ & $\checkmark$ & PCSK & $6.2\times 10^{-4}$ \\
        \hline\hline
    \end{tabular}
\end{table*}
\begin{table*}[tb!]
    \caption{\label{tab:parameters_2sigma}The 90\% confidence level intervals for the 20 model parameters extracted from the posterior distributions of runs one to four (see Tab.~\ref{tab:mcmc_setup}). The rightmost column shows the ``true'' value from the closure test.}
    \centering
    \begin{tabular}{c|c|c|c|c|c}
        \hline\hline
        Model parameter & Run 1 & Run 2 & Run 3 & Run 4 & True Value \\
        \hline
        $B_G\;[\mathrm{GeV}^{-2}]$ & $19.956_{-0.855}^{+0.878}$ & $19.861_{-1.037}^{+0.910}$ & $20.028_{-7.182}^{+4.304}$ & $19.962_{-0.606}^{+0.794}$ & $19.928$ \\
        $\alpha_{\rm shadowing}$ & $0.027_{-0.010}^{+0.011}$ & $0.032_{-0.012}^{+0.012}$ & $0.043_{-0.039}^{+0.067}$ & $0.028_{-0.010}^{+0.011}$ & $0.021$ \\
        $y_{{\rm loss},2}$ & $1.514_{-0.019}^{+0.017}$ & $1.511_{-0.022}^{+0.021}$ & $1.464_{-0.114}^{+0.115}$ & $1.514_{-0.019}^{+0.016}$ & $1.511$ \\
        $y_{{\rm loss},4}$ & $1.702_{-0.026}^{+0.028}$ & $1.714_{-0.033}^{+0.033}$ & $1.825_{-0.168}^{+0.190}$ & $1.703_{-0.025}^{+0.028}$ & $1.690$ \\
        $y_{{\rm loss},6}$ & $2.366_{-0.037}^{+0.038}$ & $2.361_{-0.045}^{+0.042}$ & $2.216_{-0.234}^{+0.250}$ & $2.368_{-0.035}^{+0.030}$ & $2.338$ \\
        $\sigma_{y_{\rm loss}}$ & $0.298_{-0.034}^{+0.021}$ & $0.295_{-0.027}^{+0.024}$ & $0.317_{-0.168}^{+0.182}$ & $0.303_{-0.019}^{+0.018}$ & $0.317$ \\
        $\alpha_{\rm rem}$ & $0.569_{-0.013}^{+0.013}$ & $0.573_{-0.013}^{+0.013}$ & $0.627_{-0.088}^{+0.092}$ & $0.570_{-0.011}^{+0.013}$ & $0.568$ \\
        $\lambda_B$ & $0.285_{-0.012}^{+0.012}$ & $0.285_{-0.014}^{+0.014}$ & $0.281_{-0.054}^{+0.056}$ & $0.286_{-0.011}^{+0.011}$ & $0.284$ \\
        $\sigma_x^{\rm string}\;[{\rm fm}]$ & $0.181_{-0.005}^{+0.004}$ & $0.179_{-0.006}^{+0.006}$ & $0.130_{-0.025}^{+0.037}$ & $0.181_{-0.005}^{+0.004}$ & $0.182$ \\
        $\sigma_\eta^{\rm string}$ & $0.120_{-0.005}^{+0.005}$ & $0.120_{-0.006}^{+0.006}$ & $0.148_{-0.031}^{+0.038}$ & $0.120_{-0.005}^{+0.005}$ & $0.122$ \\
        $\alpha_{\text{string tilt}}$ & $0.977_{-0.039}^{+0.020}$ & $0.962_{-0.053}^{+0.033}$ & $0.614_{-0.535}^{+0.354}$ & $0.982_{-0.035}^{+0.016}$ & $0.952$ \\ 
        $\alpha_{\text{preFlow}}$ & $0.081_{-0.011}^{+0.012}$ & $0.082_{-0.013}^{+0.017}$ & $0.155_{-0.100}^{+0.139}$ & $0.081_{-0.010}^{+0.013}$ & $0.086$ \\
        $\eta_0$ & $0.101_{-0.007}^{+0.006}$ & $0.100_{-0.010}^{+0.010}$ & $0.086_{-0.035}^{+0.040}$ & $0.101_{-0.006}^{+0.006}$ & $0.098$ \\
        $\eta_2$ & $0.073_{-0.008}^{+0.009}$ & $0.077_{-0.012}^{+0.013}$ & $0.130_{-0.099}^{+0.118}$ & $0.073_{-0.007}^{+0.009}$ & $0.073$ \\
        $\eta_4$ & $0.209_{-0.007}^{+0.008}$ & $0.211_{-0.010}^{+0.010}$ & $0.206_{-0.098}^{+0.079}$ & $0.209_{-0.007}^{+0.007}$ & $0.204$ \\
        $\zeta_{\rm max}$ & $0.190_{-0.005}^{+0.004}$ & $0.189_{-0.006}^{+0.006}$ & $0.182_{-0.030}^{+0.016}$ & $0.189_{-0.005}^{+0.005}$ & $0.188$ \\
        $T_{\zeta,0}\;[{\rm GeV}]$ & $0.228_{-0.003}^{+0.003}$ & $0.228_{-0.003}^{+0.003}$ & $0.226_{-0.029}^{+0.019}$ & $0.228_{-0.003}^{+0.003}$ & $0.228$ \\
        $\sigma_{\zeta,+}\;[{\rm GeV}]$ & $0.077_{-0.005}^{+0.001}$ & $0.075_{-0.003}^{+0.003}$ & $0.075_{-0.024}^{+0.023}$ & $0.077_{-0.005}^{+0.001}$ & $0.082$ \\
        $\sigma_{\zeta,-}\;[{\rm GeV}]$ & $0.041_{-0.002}^{+0.003}$ & $0.041_{-0.002}^{+0.003}$ & $0.042_{-0.021}^{+0.013}$ & $0.041_{-0.002}^{+0.003}$ & $0.041$ \\
        $e_{\rm sw}\;[{\rm GeV}/{\rm fm}^3]$ & $0.487_{-0.006}^{+0.005}$ & $0.486_{-0.007}^{+0.006}$ & $0.462_{-0.042}^{+0.033}$ & $0.487_{-0.006}^{+0.005}$ & $0.499$ \\
        \hline\hline
    \end{tabular}
\end{table*}

Comparing the first three runs in Tab.~\ref{tab:mcmc_setup} with the posterior distributions in Fig.\ref{fig:posterior_R123} gives us a good impression of the interpretation of the $\Delta$ quantity.
The values for the PCGP and PCSK emulators are two orders of magnitude smaller than the Scikit GP value, which means that they are more peaked about the true value.
We also verify that the two different MCMC approaches give consistent posterior distributions and compatible values of $\Delta$ (run 1 vs. 4).

When the HPP points are not included in the emulator training, we can see that the value of $\Delta$ increases by two orders of magnitude between run 1 and run 5.
This result is in line with the findings we had for the emulator uncertainties $\mathcal{E}$ discussed in Sec.~\ref{subsub:active_learning}.

Including the PCA and logarithmic transformations in the emulators' training does not significantly change the quality of reproducing the true parameter set in the closure test.

Table~\ref{tab:parameters_2sigma} summarizes the median values of the model parameters with 90\% confidence intervals from the obtained posterior distributions for runs 1 to 4.
Because the posterior distributions from runs 1, 2, and 4 are sharply peaked around the true parameter set, the 90\% confidence intervals are significantly narrower than those from run 3.

\section{Observable Sensitivity Analysis}
\label{sub:obs_sensitivity_analysis}

Effectively constraining model parameters through Bayesian analysis requires understanding which parameters have a strong influence on specific observables. 
In our analysis, we evaluate how parameters impact observables by calculating their first-order partial derivatives at specific points in the parameter space.
We define the response of observable $i$ w.r.t parameter $j$ by
\begin{equation}
    \mathcal{R}_{ij}\equiv \frac{\partial\ln(\bs{\mathcal{G}}_i(\bs{\theta}))}{\partial\ln(\bs{\theta}_j)}.
\end{equation}
The elements of this response matrix are unitless and show how strongly the observables depend on the parameters: positive values mean an increase and negative values mean a decrease.
To calculate the above derivative numerically we use the centered, finite difference method for first-order derivatives from the emulator predictions.

\begin{figure*}[tb!]
    \centering
    \includegraphics[width=\textwidth]{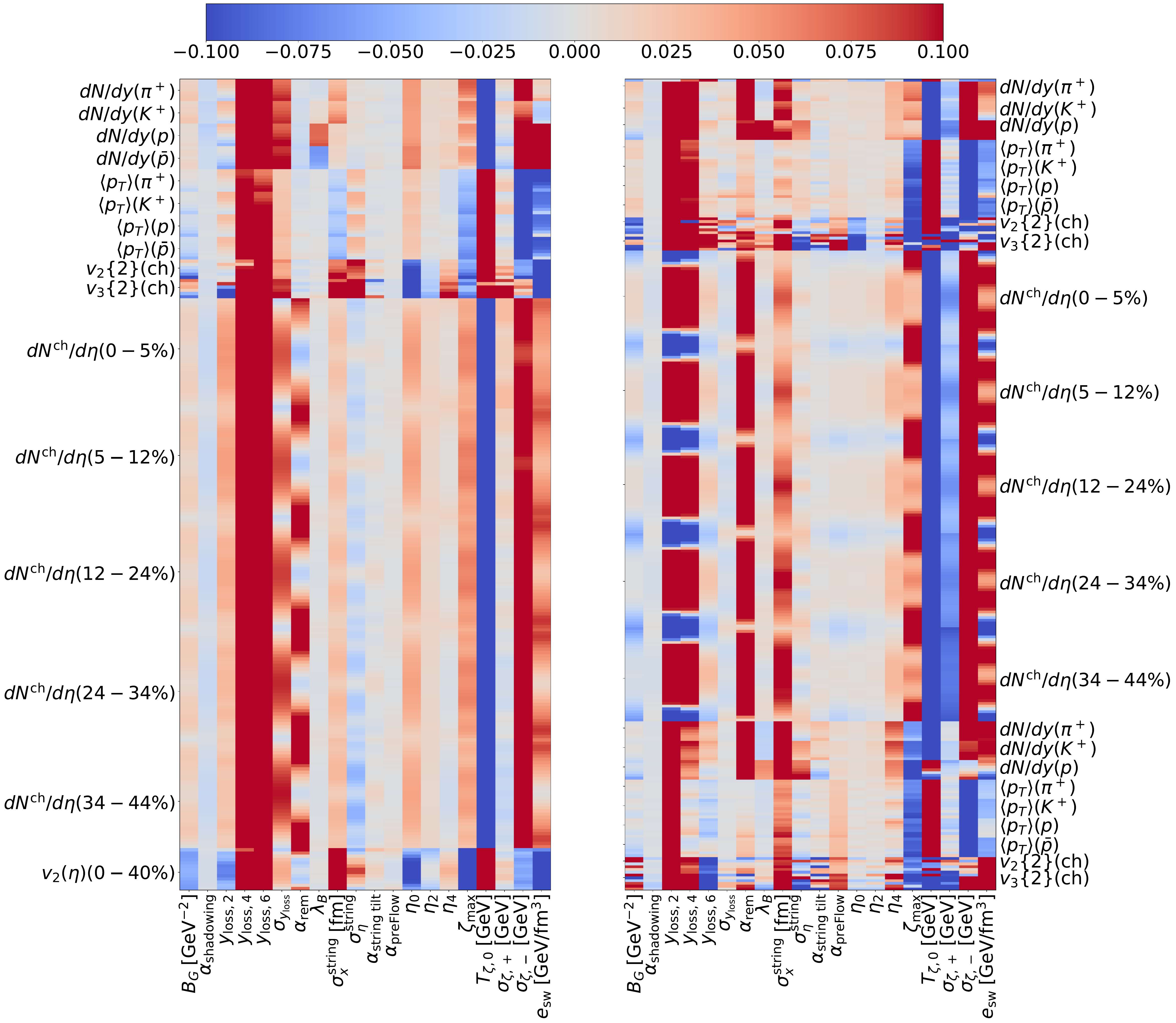}
    \caption{Response matrix $\mathcal{R}_{ij}$ at the point used in the closure test. The observables in the left figure correspond to 200 GeV collisions. The right-figure observables come from 19.6 GeV and 7.7 GeV simulations.}
    \label{fig:A_HEATMAP_SENSITIVITY_ANALYSIS}
\end{figure*}

From Fig.~\ref{fig:A_HEATMAP_SENSITIVITY_ANALYSIS}, we can identify some clear trends in how some observables are sensitive to local changes in parameter space.
As an example, we use the closure test point for the sensitivity analysis.
Let us now look at a few features in the response matrix of the observables w.r.t. the 20 parameters used in the analysis.
The rapidity distribution and mean transverse momentum distribution of $\mathrm{d}N/\mathrm{d}y$ of $\pi$, $K$ and $p$ at 200 GeV show a strong positive response to the rapidity loss parameters $y_{\mathrm{loss}, 4}$ and $y_{\mathrm{loss}, 6}$, which is natural due to the high rapidity of the incoming particles. 
On the other hand, the same observables for 7.7 GeV collisions depend strongly on the parameter $y_{\mathrm{loss}, 2}$, since the beams have a much smaller rapidity. 
We also see a strong dependence for all observables at least on some of the bulk viscosity parameters.
When the shear viscosity parameter $\eta_{0}$ increases, the anisotropic flow coefficients decrease.
A decrease in switching energy between hydrodynamics and the afterburner means a higher freeze-out temperature, which produces more particles with smaller transverse momentum.
An increase in the maximum bulk viscosity parameter leads to a decrease in the average transverse momentum and flow observables (more bulk viscosity reduces anisotropic flow, all the bulk viscosity parameters and the corresponding increase or decrease in the flow coefficients confirm this trend).

For two of the parameters, $\alpha_{\rm shadowing}$ and $\alpha_{\text{string tilt}}$, there is almost no sensitivity shown for the currently considered observables.
All the other parameters have strong sensitivities to at least one type of observable at a given energy.

In follow-up work, we will discuss the physics interpretation of the response matrix in more detail and perform the same parameter sensitivity analysis of the maximum a posteriori probability (MAP) point with the real experimental measurements at the RHIC Beam Energy Scan program.

\section{Conclusion}
\label{sec:conclusion_outlook}
In this work, we have performed a systematic comparison between three different types of open-source GP emulators -- PCGP, PCSK, and the Scikit GP -- while training them to a (3+1)D heavy-ion collision model at three collision energies.
We define a 20-dimension model parameter space and solve the Bayesian inverse problem with 544 observables related to the RHIC Beam Energy Scan program.
In addition to the difference in the emulators, we also investigated the differences in the final posterior distributions obtained from a Bayesian analysis using two different sampling methods -- MCMC and PTLMC -- in our analysis.

In the first part of this paper, we have seen that the PCGP and PCSK emulators generally give more accurate predictions and better uncertainty estimation than the Scikit GP.
We compared the standard training setup with the logarithmic transformation on the multiplicity datasets, finding, that the uncertainties are either the same or decreased in the latter case.
The difference between the approaches becomes smaller as the size of the training dataset increases.
When comparing the standard training setup with the PCA transformation on functional model inputs, $\tilde{\zeta}(T,\mu_B)$, $\tilde{\eta}(\mu_B)$ and $\langle y_{\rm loss}\rangle(y_\mathrm{init})$, we found that there is overall no significant effect in the emulation uncertainties or the metric for the uncertainty estimation of the emulators.
Even if the emulation performance does not significantly improve, this method can incorporate generic functional model inputs.
The last effect we have investigated in the emulator validation is the inclusion of the high probability posterior points, which is a simple version of an ``active learning'' ansatz with one iteration.
The inclusion of these additional points sampled from the posterior distribution significantly decreases the emulation uncertainty in the phase space where the true experimental measurements remain.

It is worth mentioning, that in terms of the prediction reliability $\mathcal{H}$, the PCGP and PCSK emulators perform better than the Scikit GP, but for some observable points, there is a significant deviation from an honest prediction ($\mathcal{H}=0$) for all emulators currently used in the field. This is certainly something users of these methods should be aware of, and it leaves room for further studies on how this changes the values of extracted parameters.

We then propagate the different emulator setups in the Bayesian inference closure test to quantify their impacts on the posterior distributions.  We observed that the PCGP and PCSK emulators provide well-defined peaked posterior distributions, while the Scikit GP result is less constrained.
The different posterior distributions are compared using the information loss quantity $\Delta$.
We obtained compatible posterior distributions from the standard MCMC and PTLMC algorithms.
Both posterior distributions are peaked around the true model parameter set. This result verified that the 544 observables used in the closure test can effectively constrain the 20-dimension model parameter space and result in the correct solution in the Bayesian inverse problem. 
We find that including the high probability posterior points in the training set significantly improves the closure test. Therefore, we recommend an active learning setup for future model emulation training.
This approach will also use computational resources more effectively and explore more details in the interesting regions in parameter space.
Lastly, we found that the logarithmic and PCA transformations do not significantly influence the accuracy of the extracted posterior distribution.

With the trained GP emulators, we can study the response matrix between the model parameters and experimental observables. Such sensitivity analysis can help us identify the most important set of observables to constrain the QGP's physical properties.

The code developed for this work is open source and can be downloaded from~\cite{hendrik_roch_2024_11186662}.

\begin{acknowledgments}
We thank Moses Chan for his helpful comments on the \texttt{surmise} Python package and the manuscript of this paper.
This work is supported in part by the U.S. Department of Energy, Office of Science, Office of Nuclear Physics, under DOE Award No.~DE-SC0021969 and DE-SC0024232. H.~R. and C.~S. were supported in part by the National Science Foundation (NSF) within the framework of the JETSCAPE collaboration (OAC-2004571).
C.S. acknowledges a DOE Office of Science Early Career Award. 
Numerical simulations presented in this work were performed at the Wayne State Grid, and we gratefully acknowledge their support.
\end{acknowledgments}

\appendix

\section{PCSK Different Training Uncertainty}
\label{app:PCSK_err}
The PCSK emulator is the only emulator in our analysis, which incorporates the statistical uncertainty at the training points.
In this case, the natural question arises, what happens if the training data is, for example, more or less precise?
We study this effect by multiplying the uncertainties of the training data by a factor of zero (label $0\sigma$) or two (label $2\sigma$) and comparing the trained emulator to the one with the original uncertainties.
\begin{figure*}[tb!]
    \centering
    \includegraphics[width=\textwidth]{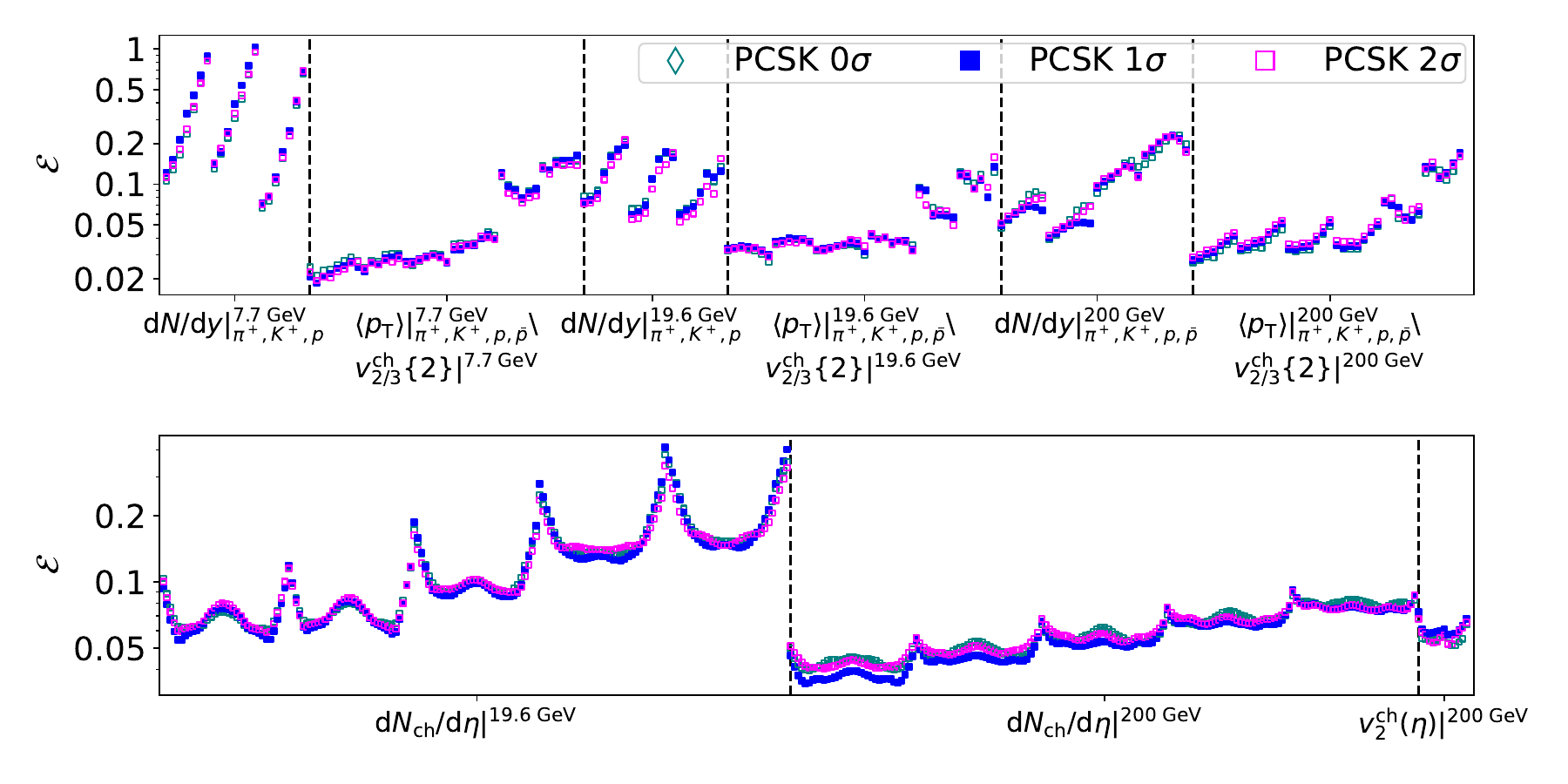}
    \caption{(Color Online) The same layout as in Fig.~\ref{fig:E_GP} but for comparing the PCSK GP emulators with different sizes of uncertainties in the training data.}
    \label{fig:E_GP_PCSK_ERR}
\end{figure*}
\begin{figure*}[tb!]
    \centering
    \includegraphics[width=\textwidth]{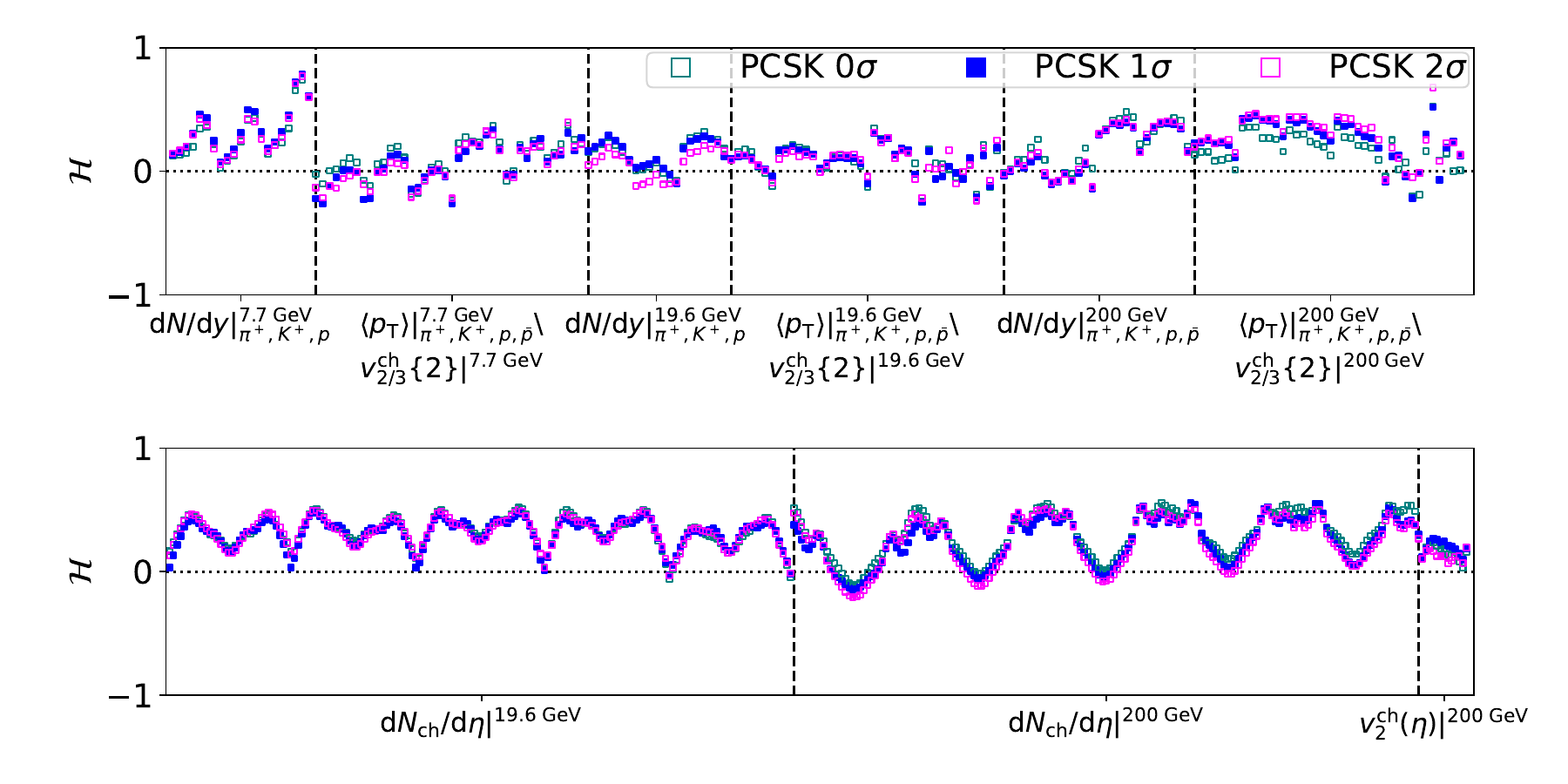}
    \caption{(Color Online) The same layout as in Fig.~\ref{fig:E_GP_PCSK_ERR} but for the metric for emulator uncertainty estimation $\mathcal{H}$.}
    \label{fig:H_GP_PCSK_ERR}
\end{figure*}
Figures~\ref{fig:E_GP_PCSK_ERR} and~\ref{fig:H_GP_PCSK_ERR} show that the RMS emulator error $\mathcal{E}$ and the metric $\mathcal{H}$ barely changes for all the observables in our analysis.
We also tried to change the uncertainty by using only half the number of events for the computation of the observables per design point (not shown).
This effect is of similar size to the one shown in Fig.~\ref{fig:E_GP_PCSK_ERR}.

Additionally, we compared the PCSK emulator with the uncertainty set to zero to the PCGP emulator, which gives similar results for the two emulators (not shown).
Small differences of the order of a few percent can be observed for $\mathcal{E}$ in the $19.6\;\mathrm{GeV}$ dataset for the multiplicities in the larger centrality classes.

\clearpage
\newpage
\bibliography{bib, non-inspire}

\end{document}